\documentclass[preprint2]{aastex}
\usepackage{rotating}

\slugcomment{To appear in ``Planets, Stars and Stellar Systems'', Vol.\ 6, Springer Publishing}

\shorttitle{GALAXY STRUCTURE}
\shortauthors{Graham}

\begin{document}
\title{
A REVIEW OF ELLIPTICAL AND DISC GALAXY STRUCTURE, AND MODERN SCALING LAWS\\
\vspace{3mm} 
{\it To appear in ``Planets, Stars and Stellar Systems'', Vol.\ 6,
  Springer Publishing. 
http://www.springer.com/astronomy/book/978-90-481-8818-5 \\
\vspace{3mm} 
{\bf Condensed / Shortened version}
} }

\author{Alister W.\ Graham\altaffilmark{1}}
\affil{Centre for Astrophysics and Supercomputing, Swinburne University
of Technology, Hawthorn, Victoria 3122, Australia.}
\altaffiltext{1}{Email: AGraham@swin.edu.au}

\begin{abstract}

A century ago, in 1911 and 1913, Plummer and then Reynolds introduced their
models to describe the radial distribution of stars in `nebulae'.
%
This article reviews the progress since then, providing both an historical
perspective and a contemporary review of the stellar structure of bulges,
discs and elliptical galaxies.
The quantification of galaxy nuclei, such as central mass deficits and excess
nuclear light, plus the structure of dark matter halos and cD galaxy
envelopes, are discussed.  Issues pertaining to spiral galaxies including
dust, bulge-to-disc ratios, bulgeless galaxies, bars and the identification of
pseudobulges are also reviewed. 
An array of modern scaling relations involving sizes, luminosities, surface
brightnesses and stellar concentrations are presented, many of which are shown
to be curved.
These `redshift zero' relations not only quantify the behavior and nature of
galaxies in the Universe today, but are the modern benchmark for evolutionary
studies of galaxies, whether based on observations, $N$-body-simulations or
semi-analytical modelling.
For example, it is shown that some of the recently discovered compact
elliptical galaxies at $1.5 < z < 2.5$ may be the bulges of modern disc
galaxies.

\end{abstract}

\keywords{
galaxy structure --- 
galaxy scaling relations --- 
galaxy elliptical --- 
galaxy spiral ---
galaxy compact ---
galaxy cD, halos ---
galaxy nuclei ---  
galaxy central mass deficits ---
galaxy excess nuclear light ---
galaxy bulge-disc ratios ---
galaxy bars ---
galaxy pseudobulges --- 
galaxy bulgeless ---
galaxy dust ---
S\'ersic model ---
core-S\'ersic model --- 
Einasto model --- 
dark matter halos.
%
}

\section{Introduction}

For the last century astronomers have been modelling the structure of
`nebulae', and here we focus on those external to the Milky Way. 
A key activity performed by many astronomers, past and present, is
the catergorisation of these galaxies (Sandage 2005) and the quantification of
their physical properties. 
How big are they? How bright are they?  What characteristics distinguish or
unite apparent subpopulations?  Answers to such questions, and the
establishment of ``scaling relations'' between two or more galactic properties
provides valuable insight into the physical mechanisms that have shaped
galaxies.

Understanding how galaxies form, increasingly through the use of simulations
and semi-analytic modelling (e.g.\ 
Cole 1991; 
White \& Frenk 1991; 
Kauffmann et al.\ 1993, 2003; 
Avila-Reese et al.\ 1998; 
Cole et al.\ 2000; 
de Lucia et al.\ 2006; 
Bower et al.\ 2006; 
Kauffmann et al.\ 2004; 
Di Matteo et al.\ 2005; 
Croton et al.\ 2006; 
Naab et al.\ 2006; 
Nipoti et al.\ 2006; 
Covington et al.\ 2011; 
Guo et al.\ 2011, and references therein), requires an accurate knowledge of galaxy
properties and scaling laws, as elucidated by Driver (2011).  
Not surprising, our knowledge of galaxies is 
best in the nearby Universe --- out to distances typically measured in megaparsecs 
rather than by redshift $z$ --- where galaxy structures can be reasonably well
resolved.  The properties of these galaxies provide the $z=0$ 
benchmark used in the calibration of galactic evolutionary studies --- both
observed and simulated.  

Popular scaling relations involving global galaxy parameters such as size,
surface brightness, luminosity and concentration are reviewed here.  As we
shall see, many bivariate distributions, which are frequently assumed to be
linear, are often only approximately so over a restricted luminosity range.
For example, it may come as a surprise for many too learn that the useful
Kormendy (1977b) relation is only the tangent to the bright arm of a
continuous but curved effective radius-(surface brightness) relation which
unifies dwarf and giant elliptical galaxies (section~\ref{Sec_Korm}).
Similarly, the Faber-Jackson (1976) relation with a slope of 4 represents the
average slope over a restricted luminosity range to what is a curved or broken
luminosity-(velocity dispersion) distribution, in which the slope is 2 rather
than 4 at lower luminosities (section~\ref{Sec_Faber}).
Knowing these trends, the bulk of which cannot be established when assuming
structural homology, i.e.\ using de Vaucouleurs' (1948) $R^{1/4}$ model, is
vital if one is to measure, model and make sense of galaxies.

This article has been structured into four main sections. 
Section 1 provides this general overview plus a further review and
introduction to galaxies on the Hubble-Jeans sequence\footnote{This review
does not encompass dwarf spheroidals, or any, galaxies fainter than $M_B
\approx -14$ mag. These galaxies can not be observed (to date) at cosmologically
interesting distances, and their increased scatter in the colour-magnitude
relation may indicate a range of galaxy types (e.g.\ Penny \& Conselice 2008, and
references therein).}.  Included are diagrams showing the location of
dynamically hot stellar systems in the mass-size and mass-density plane,
revealing
that some high-$z$ compact galaxies
have properties equivalent to the bulges of local disc galaxies.
Section~\ref{Sec_Ell} provides an historical account of how the radial
distribution of stars in elliptical galaxies have been modelled, and the iterative steps
leading to the development of the modern core-S\'ersic model
(section~\ref{Sec_cS}).  Subsections cover the S\'ersic model
(section~\ref{Sec_body}), its relation and applicability to dark matter halos
(section~\ref{Sec_dark}), partially-depleted galaxy cores
(section~\ref{Sec_def}), excess nuclear light (section~\ref{Sec_Enuc}) and
excess light at large radii in the form of halos or envelopes around giant
elliptical galaxies (section~\ref{Sec_cD}).
Section~\ref{Sec_E} presents and derives a number of elliptical galaxy scaling
relations pertaining to the main body of the galaxy.  From just two linear
relations which unite the faint and bright elliptical galaxy population
(section~\ref{Sec_Line}), a number of curved relations are derived
(section~\ref{sec_curved}).  Several broken relations, at $M_B \approx
-20.5$ mag, are additionally presented in section~\ref{Sec_bent}.
For those interested in a broader or different overview of elliptical
galaxies, some recent good reviews include Renzini (2006), Cecil \& Rose
(2007), Ciotti (2009) and Lisker (2009).
Finally, the latter third of this paper is tied up in section~\ref{Sec_Discs} 
which contains a discussion of the light 
profiles of disc galaxies and their bulge-disc decomposition (\ref{Sec_BD}). 
Also included are
subsections pertaining to dust (section~\ref{sec_dust}), the difficulties with
identifying pseudobulges (section~\ref{sec_pseudo}), potential bulgeless galaxies
(section~\ref{sec_less}) and methods to model bars (section~\ref{sec_bar}).
Throughout the article references to often overlooked discovery or pioneer
papers are provided.

\subsection{Early Beginnings} 

Looking out into the Milky Way arced across our night sky, the notion that we
are residents within a pancake-shaped galaxy seems reasonable to embrace.
Indeed, back in 1750 Thomas Wright also conjectured that we reside within a
flat layer of stars which is gravitationally bound and rotating about some
centre of mass.  However, analogous to the rings of Saturn, he entertained the
idea that the Milky Way is comprised of a large annulus of stars rotating
about a distant centre, or that we are located in a large thin spherical shell
rotating about some divine centre (one of the galactic poles).  While he had
the global geometry wrong, he was perhaps the first to speculate that faint, 
extended nebulae in the heavens are distant galaxies with their own (divine) centers.

As elucidated by Hoskin (1970), it was Immanuel Kant (1755), aware of the
elliptically-shaped nebulae observed by Maupertuis, and working from an
incomplete summary of Wright (1750) that had been published in a Hamburg
Journal\footnote{Freye Urtheile, Achtes Jahr (Hamburg, 1751), translated by
Hastie, op. cit., Appendix B.}, who effectively introduced the modern concept
of disc-like galactic distributions of stars --- mistakenly crediting Wright
for the idea.  

Using his 1.83 m ``Leviathan of Parsonstown'' metal reflector
telescope in Ireland,
Lord William Henry Parsons, the 3rd Earl of Rosse, discovered 226 New General
Catalogue\footnote{The NGC built upon the (Herschel family's) Catalog of
Nebulae and Clusters of Stars (Herschel 1864).}  (NGC: Dreyer 1888) and 7
Index Catalogue (IC: Dreyer 1895, 1908) objects (Parsons 1878). Important
among these was his detection of spiral structure in many galaxies, such as
M51 which affectionately became known as the whirlpool galaxy.

Further divisions into disc (spiral) and elliptical galaxy types followed
(e.g.\ Wolf 1908; Knox Shaw 1915; Curtis 1918; Reynolds 1920; Hubble
1926)\footnote{Reynolds (1927) called Hubble's attention to pre-existing and
partly-similar galaxy classification schemes which were not cited.}  and
Shapley \& Swope (1924) and Shapley (1928) successfully identified our own
Galaxy's (gravitational) center towards the constellation Sagittarius (see
also Seares 1928).

With the discovery that our Universe contains Doppler shifted 'nebulae' that
are expanding away from us (de Sitter 1917; Slipher 1917; see also Friedmann
1922, Lundmark 1924 and the reviews by Kragh \& Smith 2003 and Shaviv 2011), in accord with 
a redshift-distance relation (Lemaitre 1927, Robertson 1928, Humasson 1929, 
Hubble 1929)\footnote{It is of interest to note that Hubble (1934, 1936b,
1937) was actually cautious to accept that the redshifts corresponded to real 
velocities and thus an expanding Universe as first suggested by others. 
He used the term ``apparent velocity'' to flag his skepticism. 
In point of fact, Hubble \& Tolman (1935) wrote that the data is ``not yet 
sufficient to permit a decision between recessional or other causes for the
red-shift''.} 
--- i.e.\ awareness that some of the ``nebuale'' are external objects to our galaxy --- came
increased efforts to catergorise and organise these different types of
``galaxy''.
As noted by Sandage (2004, 2005), Sir James Jeans (1928) was the first to
present the (tuning fork)-shaped diagram that encapsulated Hubble's (1926)
early-to-late type galaxy sequence, a sequence which had been inspired in part
by Jeans (1919) and later popularised by Hubble (1936a; see Block et al.\
2004).  Quantifying the physical properties of galaxies along this sequence,
with increasing accuracy and level of detail, has occupied many astronomers
since.
Indeed, this review addresses aspects related to the radial concentration of
stars in the elliptical and disc galaxies which effectively define the
Hubble-Jeans sequence.  Irregular galaxies are not discussed here.

\subsection{The modern galaxy}

\begin{figure*}
\includegraphics[angle=-90,scale=0.35]{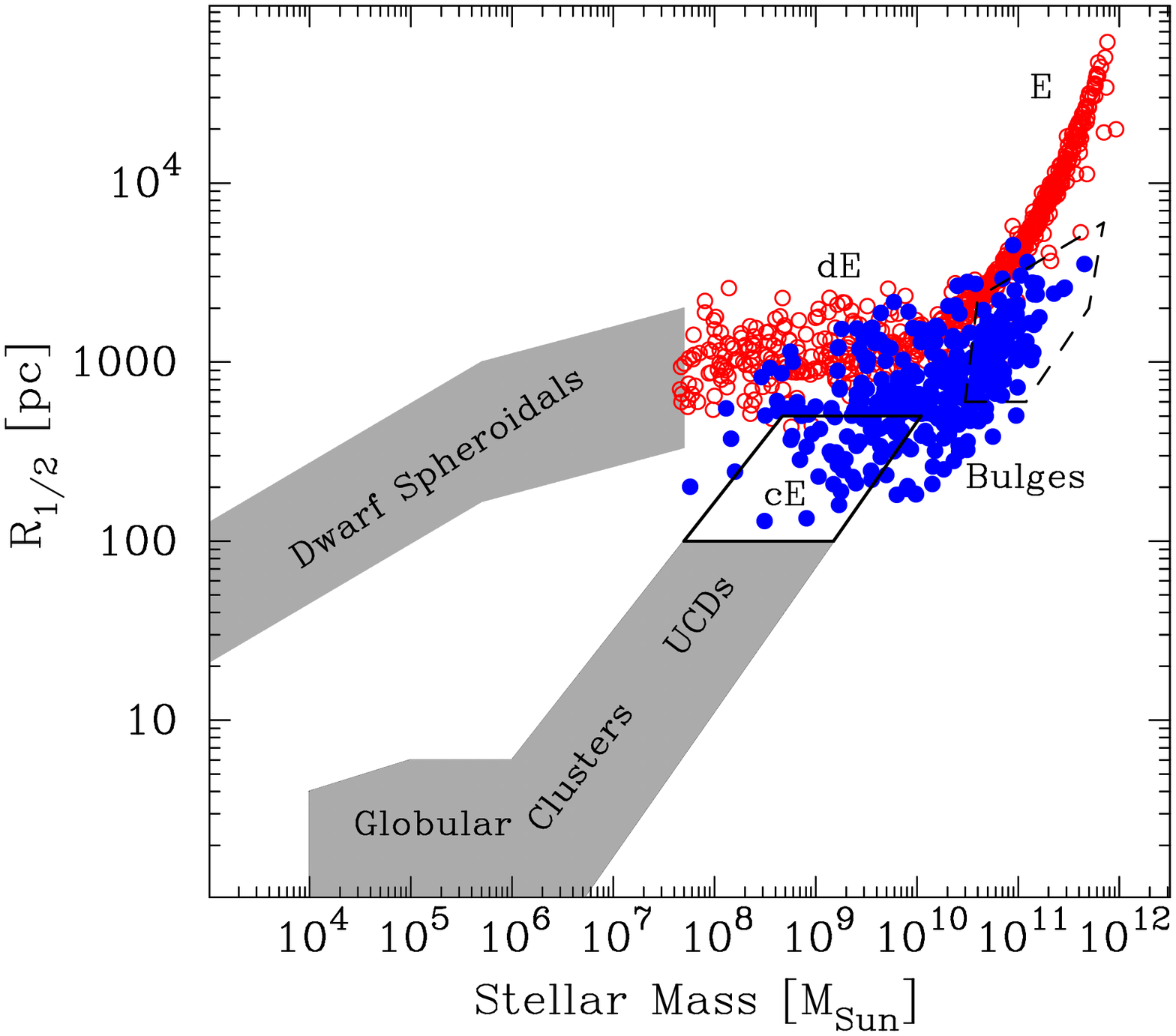}
\hspace{10mm}
\includegraphics[angle=-90,scale=0.35]{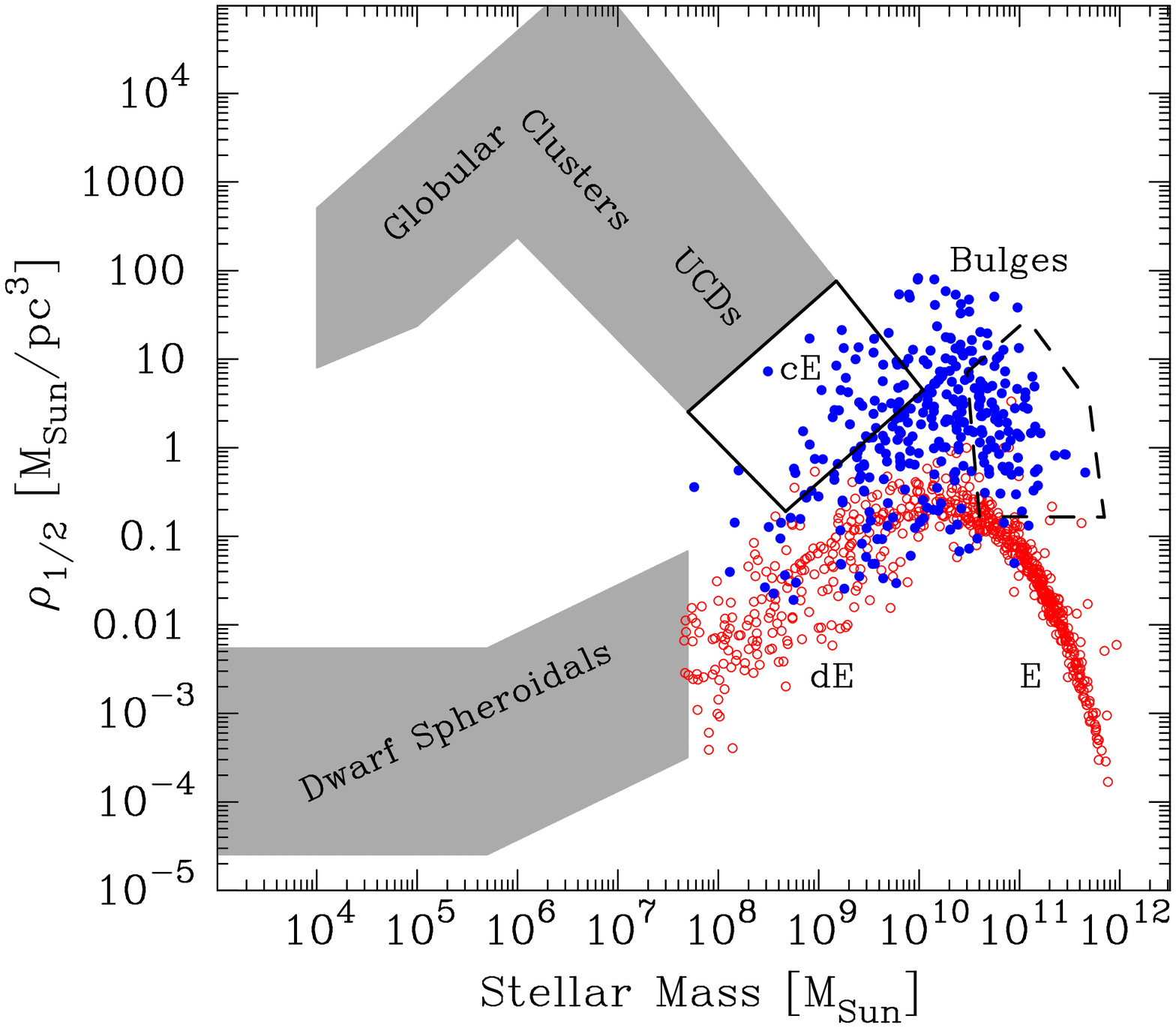}
\caption{
Left panel: The radius containing half of each object's light, $R_{1/2}$ 
(as seen in projection on the sky), 
is plotted against each object's stellar mass. Open 
circles: dwarf elliptical (dE) and ordinary elliptical (E) galaxies from
Binggeli \& Jerjen (1998), Caon et al.\ (1993), D'Onofrio et al.\ (1994) and
Forbes et al.\ (2008). Filled circles: Bulges of disc galaxies from Graham \&
Worley (2008).  Shaded regions adapted from Binggeli et al.\ (1984, their
figure~7), Dabringhausen et al.\ (2008, their
figure~2), Forbes et al.\ (2008, their figure~7), Misgeld \& Hilker (2011,
their figure~1). The location of the so-called ``compact elliptical'' (cE)
galaxies is shown by the rhombus overlapping with small bulges. 
The location of dense, compact, $z=1.5$ galaxies, as indicated by
Damjanov et al.\ (2009, their figure~5), is denoted by the dashed boundary 
overlapping with luminous bulges.
Right panel: Stellar mass density within the volume containing half each object's 
light, $rho_{1/2}$, versus stellar mass.  The radius of this volume was taken
to equal 4/3 $\times R_{1/2}$ (Ciotti 1991; Wolf et al.\ 2010). 
}
\label{Fig1}
\end{figure*}

For reasons that will become apparent, this review uses the galaxy notation of
Alan Sandage and Bruno Binggeli, in which dwarf elliptical (dE) galaxies are
the faint extension of ordinary and luminous elliptical (E) galaxies, and the
dwarf spheroidal (dSph) galaxies --- prevalent in our Local Group (Grebel
2001) --- are found at magnitudes fainter than $M_B \approx -13$ to $-$14 mag
($\approx 10^8 M_{\odot}$ in stellar mass; see Figure~\ref{Fig1}a).  Figure~\ref{Fig1}a
reveals a second branch of elliptically-shaped object stretching from the
bulges of disc galaxies and compact elliptical (cE) galaxies to ultra compact
dwarf (UCD) objects (Hilker et al.\ 1999; Drinkwater et al.\ 2000; Norris \&
Kannappan 2011 and references therein). 
A {\it possible} connection is based upon
the stripping of a disc galaxy's outer disc to form a cE galaxy (Nieto 1990;
Bekki et al.\ 2001b; Graham 2002; Chilingarian et al.\ 2009) and through
greater stripping of the bulge to form a UCD (Zinnecker et al.\ 1988; Freeman
1990; Bassino et al.\ 1994; Bekki 2001a).  It is thought that nucleated dwarf
elliptical galaxies may also experience this stripping process,
giving rise to UCDs.

While the identification of local spiral galaxies is relatively free from
debate, the situation is not so clear in regard to elliptically-shaped
galaxies.  The discovery of UCDs, which have sizes and fluxes intermediate
between those of galaxies and (i) the nuclear star clusters found at the
centres of galaxies and (ii) globular clusters (GCs: e.g.\ Ha{\c s}egan et
al.\ 2005; Brodie \& Strader 2006), led Forbes \& Kroupa (2011) to try and
provide a modern definition for what is a galaxy (see also Tollerud et al.\
2011).  Only a few years ago there was something of a divide between GCs and
UCDs --- all of which had sizes less than $\sim$30 pc --- and galaxies with
sizes greater than ~120 pc (Gilmore et al.\ 2007).  However, as we have
steadily increased our celestial inventory, objects of an intermediate nature
have been found (e.g.\ Ma et al.\ 2007, their Table~3), raising the question
asked by Forbes \& Kroupa for which, perhaps not surprisingly, no clear answer
has yet emerged.  While those authors explored the notion of a division by,
among other properties, size and luminosity, they did not discuss how the
density varies.  As an addendum of sorts to Forbes \& Kroupa (2011), the
density of elliptically-shaped objects is presented here in
Figure~\ref{Fig1}b.  This is also done to allow the author to wave the
following flag.

Apparent in Figure~\ref{Fig1}b, but apparently not well recognised within the
community, is that the bulges of disc galaxies can be much denser than elliptical galaxies.  If the
common idea of galaxy growth via the accretion of a disc, perhaps from
cold-mode accretion streams, around a pre-existing spheroid is correct (e.g.\
Navarro \& Benz 1991;
Steinmetz \& Navarro 2002; Birnboim \& Dekel 2003; 
see also Conselice et al.\ 2011 and Pichon et al.\ 2011), then one should
expect to find dense spheroids at high-$z$ with $10^{10}$--$10^{11} M_{\odot}$
of stellar material, possibly surrounded by a faint (exponential) disc which
is under development.
It is noted here that the dense, compact early-type galaxies recently 
found at redshifts of 1.4-2.5 (Daddi et al.\ 2005; Trujillo et al.\ 2006)
display substantial overlap with the location of present day bulges in
Figure~\ref{Fig1}a, and that the merger scenarios for 
converting these compact high-$z$ galaxies into today's elliptical galaxies are not 
without problems (e.g.\ Nipoti et al.\ 2009; Nair et al.\ 2011). 
It is also noted that well-developed discs and disc galaxies are rare at the 
redshifts where these compact objects have been observed alongside
normal-sized elliptical galaxies. 
Before trying to understand galaxy structure at high-redshift, and galaxy
evolution --- themes not detailed in this review --- it is important to 
first appreciate galaxy structures at $z=0$ where observations are easier and local
benchmark scaling relations have been established.

\section{Elliptical Galaxy Light Profiles}\label{Sec_Ell}

Over the years a number of mathematical functions have been used to represent
the radial distribution of stellar light in elliptical galaxies, i.e.\ their
light profiles.  Before getting to de Vaucouleurs' $R^{1/4}$ model in the
following paragraph, it seems apt to first quickly mention some early
competitors.
Although Plummer's (1911) internal-density model was developed for the nebulae
which became know as globular
clusters, because of its simplicity it is still used today by some researchers
to simulate elliptical galaxies, even though, it should be noted, no modern
observers use this model to describe the radial distribution of light in
elliptical galaxies. 
Reynold's (1913) surface-density model, sometimes referred to as Hubble's
(1930) model or the Reynold-Hubble model,
was used to describe the nebula which became known as elliptical galaxies. 
It has an infinite mass and is also no longer used by observers today.  The
modified Hubble model (Rood et al.\ 1972), which also has an infinite mass,
is also still sometimes used by simulators, even though, again, observers do
not use this model anymore.  Oemler's (1976) exponentially-truncated Hubble model,
known as the Oemler-Hubble model, is also not used to represent the observed
stellar distribution in elliptical galaxies because it too, like its
predecessors, was simply an approximation applicable over a limited radial
range, as noted by King (1978).
It is interesting to note that up until the 1980s, departures at large radii
from the Reynold's model were attributed to tidal-stripping by external
gravitational potentials. That is, for three quarters of a century, Reynold's
model --- originally developed from low-quality data for one galaxy --- was
generally thought to describe the original, undisturbed stellar distribution in elliptical
galaxies.

de Vaucouleurs' (1948, 1953) $R^{1/4}$ surface-density model had traction for
many years, in part due to de Vaucouleurs (1959) arguing that it fits better than
the Reynold's model used by Hubble, --- a point re-iterated by Kormendy
(1977a) and others --- and the revelation that it fits the radially-extended
data for NGC~3379 exceedingly well (de Vaucouleurs \& Capaccioli 1979).
Hodge (1961a,b) had however revealed that de Vaucouleurs' model was
inadequate to describe faint elliptical galaxies 
and Hodge (1963, 1964), in addition to King (1962)\footnote{King (1962) also
  noted that his model failed to fit the inner region of bright elliptical
  galaxies.}, noted that the 3-parameter King model, with its flatter
 inner profile and steeper decline at large radii, did a better job.
For a time, King's (1962, 1966) model became popular for describing the light
distribution in faint elliptical galaxies, at least until the exponential
model
 --- also used for the discs of spiral galaxies --- was noted to provide
a good description of some dwarf elliptical galaxies (Hodge 1971; Faber \& Lin
1983; Binggeli et al.\ 1984) and that these galaxies need not have experienced 
any tidal truncation (a prescription of the King model with its tidal radius parameter). 
Lauer (1984, 1985) additionally showed that King's modified isothermal model,
with its flat inner core, was inadequate to describe the deconvolved
light-profiles of ordinary elliptical galaxies with ``cores'', i.e.\ galaxies
whose inner light profile displays a nearly flat core.  King's model does
however remain extremely useful for studies of star clusters, globular
clusters\footnote{The Wilson (1975) and Elson (1987) model are also useful
 for describing globular clusters.}, dwarf spheroidal galaxies and galactic satellites which, unlike
ordinary elliptical galaxies, can have flat cores in their inner surface
brightness profile.

Today, the model of choice for describing nearby (and distant) dwarf and
ordinary elliptical galaxies is S\'ersic's (1963) generalisation of de 
Vaucouleurs' $R^{1/4}$ model to give the $R^{1/n}$ surface-density model
(section~\ref{Sec_body}). 
This model reproduces the exponential model when $n=1$ and de Vaucouleurs'
model when $n=4$; it can thus describe the main body of faint and luminous
elliptical galaxies.  The key advantage that this model has is (i) its ability to
describe the observed stellar distributions that have a range of central
concentrations (known to exist since at least Reaves 1956) 
and (ii) it provides a very good description of the data over (almost) 
the entire radial extent.  Indeed, departures in the light profile from a
well-fit S\'ersic's model invariably signal the presence of additional
features or components, rather than any failing of the model. 
Expanding upon the S\'ersic model, the 
core-S\'ersic model (section~\ref{Sec_cS}) is nowadays used to quantify those galaxies
with ``cores''. 

Although referring to the King model, the following quote from King (1966)
seems particularly insightful ``... de Vaucouleurs' law appears to refer to a
particular central concentration and should be appropriate only for galaxy
profiles that have that concentration.''  While noted by others, such as
Oemler (1976), Capaccioli (1985), 
Michard (1985) and Schombert (1986), some three decades elapsed before the 
relevance of King's remark to elliptical galaxies re-surfaced --- albeit
slowly at first --- in the 1990s.  Indeed, de Vaucouleurs' useful, albeit
limited, $R^{1/4}$ model was referred to as a ``law'' for nearly half a 
century.
However we are now more keenly aware that (even normal) elliptical galaxies
possess a range of central concentrations: concentrations which are well
quantified in terms of the exponent $n$ in S\'ersic's $R^{1/n}$ model
(see Trujillo, Graham \& Caon 2001; Graham, Trujillo \& Caon 2001).
Although, it should be confessed that one can still encounter papers which use
$R^{1/4}$ model parameters alongside some model-independent measure of galaxy
concentration, unaware of the inconsistency arising from the fact that every
$R^{1/4}$ model actually has exactly the same level of concentration.

Before introducing the equation for S\'ersic's model in the following section, it is
pointed out that in addition to modelling what can be considered the main body
of the galaxy, one can also find excess stellar light at (i) small radii in
the form of nuclear (i.e. centrally located) discs and dense nuclear star
clusters (section~\ref{Sec_Enuc}) and also at (ii) large radii in the form of halos or
envelopes in cD and central cluster galaxies (section~\ref{Sec_cD}).  
As briefly noted above, deficits of 
stellar flux at the cores of massive galaxies are also observed, and a model
to quantify these stellar distribution, relative to the outer non-depleted
light profile, is described in section~\ref{Sec_cS}.
While non-symmetrical components in elliptical galaxies can also exist, they
are not addressed here given the focus on well-structured systems.  Somewhat
random, non-symmetrical components may be a sign of a disturbed morphology
(see E.Barton's Chapter in this volume), of on-going non-uniform star formation
(see S.Boissier's Chapter in this volume) or gravitationally-induced tidal
features from external forces.

\subsection{S\'ersic's model}\label{Sec_body}

Jos\'e S\'ersic's (1963, 1968) $R^{1/n}$ model, which was introduced in Spanish, 
describes how the projected
surface-intensity $I$ varies with the projected radius $R$, such that
\begin{equation}
I(R)=I_{\rm e}\exp\left\{ -b_n\left[\left( \frac{R}{R_{\rm e}}\right) ^{1/n}
  -1\right]\right\} 
\label{Eq_Ser}
\end{equation}
and $I_{\rm e}$ is the intensity at the `effective' radius $R_{\rm e}$
that encloses half of the total light from the model (Ciotti 1991; Caon
et al.\ 1993). 
The term $b_n$ ($\approx 1.9992n-0.3271$ for $0.5 < n < 10$, Capaccioli 1989) 
is not a parameter but is instead 
dependent on the third model parameter, $n$, that describes the
shape, i.e.\ the concentration, of the light profile.\footnote{Ellipticity
gradients 
result in a different S\'ersic index for the major- and minor-axis, as noted
by Caon et al.\ (1993) and later quantified by Ferrari et al.\ (2004).}
The exact value of $b_n$ is obtained by solving the equation
$\Gamma(2n)=2\gamma (2n,b_n)$ where $\gamma $(2n,x) is the incomplete gamma
function and $\Gamma $ is the (complete) gamma function (Ciotti 1991).  Useful
S\'ersic related expressions have been presented in Ciotti (1991), Simonneau
\& Prada (2004) and Ciotti \& Bertin (1999), while Graham \& Driver (2005)
provide a detailed review of S\'ersic's model plus associated quantities and
references to pioneers of this model.

\begin{figure*}
\includegraphics[angle=-90,scale=0.7]{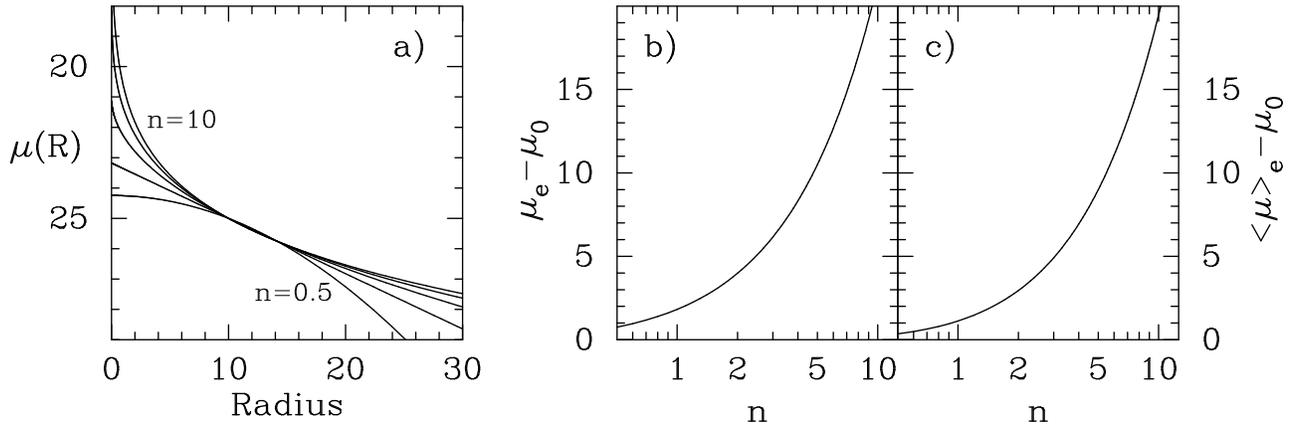}
\caption{
Left panel: S\'ersic's $R^{1/n}$ model (equation~\ref{Eq_Ser}) for indices $n=0.5, 1, 2,
4$ and 10.  The effective radius ($R_{\rm e}$) and surface brightness 
($\mu_{\rm e} = -2.5\log\, I_{\rm e}$) have been arbitrarily set to 10 and 25.
Right panel: Difference between the various surface brightness terms discussed
in the text. 
}
\label{Fig_Ser}
\end{figure*}

The relation between the effective surface brightness ($\mu_{\rm e} = -2.5\log
I_{\rm e}$) and the central surface brightness ($\mu_0 = -2.5\log
I_0$) is given by the expression 
\begin{equation}
\mu_e = \mu_0 + 1.086b, 
\label{Eq_mue-mu0}
\end{equation}
where we have dropped the subscript $n$ from the term $b_n$ for simplicity, 
while
\begin{equation}
\langle \mu \rangle _{\rm e} = \mu_{\rm e} -2.5\log[e^{b}n\Gamma(2n)/b^{2n}]
\label{Eq_mue-mue}
\end{equation}
gives the difference between the effective surface brightness and the 
mean effective surface brightness ($\langle \mu \rangle _{\rm e}$) within
$R_{\rm e}$.  Figure~\ref{Fig_Ser} shows the behaviour of the S\'ersic model. 

Uniting CCD data with wide and deep photographic images, Caon et al.\ (1990,
1993, 1994) revealed that the S\'ersic $R^{1/n}$ model provided a remarkably good
description to the stellar distribution over a large radial range, down to
surface brightnesses of $\sim$28 $B$-mag arcsec$^{-2}$, for the early-type
galaxies brighter than $M_B=-18$ mag in the Virgo cluster.  
This work was in essence an expansion of de Vaucouleurs
\& Capaccioli's (1979) study of NGC~3379 which is very well-fit with
$n=4$.  Different galaxies were discovered by Caon et al.\ (1993) to be equally well fit, but
required different values of $n$ (see also Bertin et al.\ 2002). 

Importantly, Caon et al.\ (1993) additionally showed that a
correlation existed between stellar concentration, i.e.\ the S\'ersic index
$n$, and (model-independent) galaxy size that was not due to parameter
coupling in the S\'ersic model (see also Trujillo et al.\ 2001, their
section~2).  One of the commonly overlooked implications of this result is
that $R^{1/4}$, and similarly Petrosian (1976), magnitudes, sizes and surface brightnesses
are systematically in error as a function of galaxy concentration (Graham et
al.\ 2005; Hill et al.\ 2011). 
That is, application of a model which 
fails to adequately capture the range of stellar distributions will result 
in parameters which are systematically biased as a function of galaxy
mass.  For example, fitting an
$R^{1/4}$ model to elliptical galaxies which are actually described by an
$R^{1/n}$ model with $n$ less than and greater than 4 will yield sizes and
luminosities which are, respectively, greater than and less than the true
value (e.g.\ Binggeli \& Cameron 1991; 
Trujillo et al.\ 2001; Brown et al.\ 2003).  Similarly, fitting
an exponential model to bulges that are best described by an $R^{1/n}$ model
with $n$ less than and greater than 1 will yield sizes and luminosities which
are, respectively, greater than and less than the true value (e.g.\ Graham
2001).  
Obviously one does not want to fine tune their galaxy simulations to match
scaling relations that contain systematic biases due to poor measurements, and
observers are therefore busy fitting $R^{1/n}$ models these days. 

A good approximation to the internal-density profile associated with
S\'ersic's model, i.e.\ with its deprojection, was introduced by Prugniel \&
Simien (1997).  Useful expressions for the dynamics, gravitational potential
and forces of this model have been developed by Trujillo et al.\ (2002), 
Terzi\'c \& Graham (2005) and Terzi\'c \& Sprague (2007).
Somewhat more complex than the early light-profile 
models, such expressions can, importantly, accommodate a 
range of concentrations, rather than only varying one scale radius and one
scale density. 
Such a model is vital if one wishes to properly simulate and understand the mass spectrum
of elliptical galaxies, whose S\'ersic index $n$ increases with stellar 
mass.  While Graham \& Driver's (2005) review stated that 
``No attempt has been made here to show the numerous scientific advances
engendered via application of the $R^{1/n}$ model'', this article 
reveals how S\'ersic's model, and the core-S\'ersic model 0
(subsection~\ref{Sec_cS}), 
have become key in unifying and understanding the galaxies around us.

Like the majority of surface- and internal-density models from the last 
century, the S\'ersic function is an empirical model created to match data
rather than developed from theory, and as such we should be cautious before
calling it a law. 
Attempts to find a physical explanation for de Vaucouleurs' model yielded
results which helped to keep it in vogue.  
Dissipational models have long been touted for producing $R^{1/4}$ profiles
(e.g.\ Larson 1969, 1974), and in the 1980s papers based on
dissipationless N-body simulations of a cold clumpy collapse or the merger of
disc galaxies also claimed to finally produce $R^{1/4}$ (and also Reynold) profiles (e.g.\
van Albada 1982; McGlynn 1984; Carlberg et al.\ 1986; Barnes 1988).
However a closer inspection reveals clear departures from the $R^{1/4}$
profile, with the simulated profiles better described by an $R^{1/n}$ model
with $n<4$.  Obviously their inability (or perhaps lack of desire, although
see Farouki, Shapiro \& Duncan 1983 whose non-homologous merger remnants were
initially criticised by $R^{1/4}$ aficionados) to create the range of stellar
concentrations now observed in elliptical galaxies highlights a limitation of
these early works.  Nonetheless, these pioneering studies have led to N-body
simulations by Nipoti et al.\ (2006) and Aceves et al.\ (2006) --- and Farouki
et al.\ (1983), whose results with a smaller force softening appeared years
ahead of their time --- have now recovered a range of S\'ersic profile shapes
for gravitational collapses in a dark matter halo and for disc galaxy mergers,
respectively.

Given the empirical nature of S\'ersic's $R^{1/n}$ model, 
Hjorth \& Madsen (1995) revealed how 
dissipationless merging and violent relaxation
provided a physical explanation for the departure from the homologous 
$R^{1/4}$ model.  Other works have explained how the quasi-constant specific entropy
associated with the post violent-relaxation stage of elliptical galaxies
results in the observed mass-dependent range of stellar concentrations in
elliptical galaxies (Gerbal et al.\ 1997; Lima Neto 1999; M\'arquez et al.\ 2001). 

It does not seem too unreasonable to speculate that elliptical galaxies,
whether built by near-monolithic collapse, collisions of disc galaxies, wet or
dry mergers, appear to eventually experience the same force(s) of nature that
results in their radial stellar distribution depending on the total stellar mass.
That is, it may not matter how the mass was accumulated into an elliptical
galaxy, once it becomes a dynamically-heated, bound stellar-system, it appears
to eventually obey certain universal scaling relations (see section~\ref{Sec_E}).

It is interesting to note that 
S\'ersic actually introduced his model as a way to parameterise disc galaxies
which he thought were comprised of differing ratios of a disc plus an
$R^{1/4}$-bulge.  His model was not initially intended to fit elliptical galaxies,
and as such it did not immediately threaten de Vaucouleurs' model.
Credit for popularising the use of S\'ersic's $R^{1/n}$ model for
approximating not only lenticular\footnote{This term was introduced by Knox Shaw (1915) and
  Reynolds (1920) in their galaxy classification scheme.} 
bulge$+$disc galaxies but for describing pure elliptical galaxies 
resides largely with Massimo Capaccioli 
(e.g.\ Capaccioli 1985, 1987, 1989; Caon et al.\ 1993; D'Onofrio et al.\
1994)\footnote{It is worth noting that D'Onofrio (2001) re-modelled the Virgo
and Fornax 2-component lenticular galaxies with an $R^{1/n}$-bulge plus an
exponential disc (see section~\ref{Sec_Discs}.}.  However, Davies et al.\ (1988) 
had also introduced this model for dwarf elliptical galaxies, while Sparks (1988)
developed an early Gaussian seeing correction for this model, and Ciotti 
(1991) developed a number of associated expressions such as the velocity
dispersion profile and a distribution function.  The important quantification that
Capaccioli and others provided is how the radial distribution of stars in 
elliptical galaxies, i.e.\ their concentration, varies with the size, luminosity and
thus the mass of the elliptical galaxy (see also Cellone, Forte, \& Geisler
1994; Vennik \& Richter 1994; Young \& Currie 1994, 1995; Graham et al.\ 1996;
Karachentseva et al.\ 1996, Vennik et al.\ 1996).  As we shall see in 
this article, the implications of this breakthrough have been 
dramatic, unifying what had previously been considered two distinct species of
galaxy, namely dwarf and ordinary elliptical galaxies, previously thought to
be described by an exponential and $R^{1/4}$ model, respectively.

\subsubsection{Dark Matter Halos}\label{Sec_dark}

This review would be somewhat incomplete without a few words regarding the
connection between S\'ersic's model and (simulated) dark matter halos. 
While modified theories of gravity may yet make dark matter redundant at some 
level, 
it is intriguing to note that the Prugniel-Simien (1997) internal-density
model, developed to approximate the deprojected form of S\'ersic's $R^{1/n}$
model, additionally provides a very good representation of the
internal-density profiles of simulated dark matter halos. Merritt et al.\
(2006) revealed that it actually provides a better description than not only
the Navarro, Frenk \& White (1997) model but even a generalised NFW model with
an arbitrary inner profile slope $\gamma$.

S\'ersic's former student, Navarrro, independently applied 
S\'ersic's surface-density model to the internal-density profiles of
simulated dark matter halos (Navarro et al.\ 2004; Merritt et al.\ 2005).  
Jaan Einasto (1965) had previously 
developed this same function as S\'ersic to describe the internal-density
profiles of galaxies.  Rather than a universal profile shape, as advocated by
Navarro et al.\ (1997), a range of simulated dark matter density profile
shapes is now known to vary with the dark matter halo mass (Avila-Reese et
al.\ 1999; Jing \&
Suto 2000; Merritt et al.\ 2005; Del Popolo 2010 and references therein).  A
number of useful expressions related to this ``Einasto model'', which has the
same functional form as S\'ersic's model but is applied to the internal rather
than projected density profile, can be found in Cardone et al.\ (2005), Mamon
\& {\L}okas (2005) and Graham et al.\ (2006).

An apparent ``bulge-halo conspiracy'' between the radial distribution of
stellar mass and dark matter (after modification by baryons) has 
arisen in recent years, such that elliptical galaxies reportedly have {\it
total} internal-density profiles $\rho(r)$ described by power-laws (Bertin \&
Stiavelli 1993; Kochanek 1995; Gavazzi et al.\ 2007; Buote \& Humphrey 2011).
These power-laws were originally claimed to be close to isothermal, such that
$\rho(r) \propto r^{-2}$ (Koopmans et al.\ 2006, 2009; Gavazzi et al.\ 2007).
Recent work now emphasises that only the sample average profile slope is
close to $-2$, and that a trend in slope with galaxy size exists 
(Humphrey \& Boute 2010; Auger et al.\ 2010). 
This is a developing field and it is worth noting that the analyses have been
confined to massive galaxies with velocity dispersions greater than $\sim$175
km s$^{-1}$, and thus with S\'ersic indices $n \gtrsim 4$ (Graham,
Trujillo \& Caon 2001).  While the light profile shape changes dramatically as
the S\'ersic index $n$ increases from $\sim$1 to $\sim$4, there is not such an
obvious change in light profile shape from $\sim$4 to higher values of $n$
(see Figure~\ref{Fig_Ser}).  The apparent isothermal profiles of elliptical
galaxies, and the alleged
``bulge-halo conspiracy'', may turn out to be a by-product of sample selection
(i.e.\ choosing galaxies which have approximately the same structure). 
It would be interesting to expand the Sloan Lens ACS (SLACS) Survey (Bolton et
al.\ 2006) to a greater range than only bright early-type galaxies that are
approximately well fit with an $R^{1/4}$ model, and to go beyond the use of
simple power-laws to describe the total mass density profile once the data
allows this. Two component mass models (Prugniel-Simien$+$Einasto) to a range of
galaxy masses await. 
Claims that ``early-type galaxies are structurally close to homologous'' may
therefore be premature, as was the case for the distribution of stellar light in elliptical
galaxies while the $R^{1/4}$ model was thought to be a law.

\subsection{The core-S\'ersic model}\label{Sec_cS}

The centres of luminous galaxies have long been known to possess ``cores'', 
such that the surface-density profile flattens at the center (e.g.\ King \&
Minkowski 1966), and 
King \& Minkowski (1972) remarked on the inability of the Reynold's and
de Vaucouleurs' model to match these flattened cores in giant galaxies. 
Although King (1978) identified a number of galaxies thought to be well
described by his model, using seeing-deconvolved ground-based images, Lauer
(1983, 1984, 1985) analysed 14 galaxies with 'cores', ranging from 1.5--5.0
arcseconds in radius, revealing that they, like M87 (Young et al.\ 1978;
Duncan \& Wheeler 1980; Binney \& Mamon 1982), were not exactly described by
the King model which had a completely flat core.  Similar conclusions, that
cores existed but that they do not have flat inner surface brightness
profiles, were also reported by Kormendy (1982, 1985a), creating the need for a
new model to describe the stellar distribution in galaxies.

Nearly a decade later, the Hubble Space Telescope was flying and offered
factors of a few improvement over the best image resolution achievable from the
ground at that time.  Not surprisingly, astronomers explored the centres of
galaxies.  In an effort to quantify these nuclear regions, after the
abandonment of the King model and the lack of a flattened core in the $R^{1/4}$ model,
Crane et al.\ (1993), Ferrarese et al.\ (1994), Forbes et al.\ (1994) and
Jaffe et al.\ (1994) used a double power-law model to describe the inner light
profiles of large galaxies.  
Grillmair et al.\ (1994), Kormendy et al.\ (1994) and Lauer et al.\ 
(1995) also adopted a double-power-law model but one with an additional, fifth,
parameter to control the sharpness of the transition.  Their model, which they
dubbed the ``Nuker law'' for describing the nuclear regions of galaxies (after
excluding any apparent excess light), has the same functional form as the
double power-law model presented by Hernquist (1990, his 
equation~43) to describe the internal-density of galaxies (Zhao 1996).

However, as noted by the above authors, these double power-law models were
never intended to describe the entire radial extent of a galaxy's stellar
distribution, and they provided no connection with the outer ($R^{1/4}$-like)
radial profile.  This disconnection turned out to be their down fall.  Due to
the curved nature of the outer light profiles beyond the core, which were being
fitted by the double power-law model's outer power-law, the five parameters of the Nuker model
systematically changed as the fitted radial extent changed.  This was first 
illustrated in a number of diagrams by Graham et al.\ (2003) who revealed that
none of the Nuker model parameters were robust, and as such they could not
provide meaningful physical quantities.  For example, Trujillo et al.\ (2004)
reported that the Nuker-derived core-radii were typically double, and up to a
factor of five times larger, than the radius where the inner power-law core
broke away from the outer $R^{1/4}$-like profile --- a result reiterated by
Dullo \& Graham (2011, in prep.).
An additional problem was that these ``break radii'' were being identified in the 
so-called ``power-law'' galaxies that showed no evidence of a downward
departure and flattening from the inward extrapolation of the outer
$R^{1/4}$-like profile. This situation arose because of the curved nature of
what were actually S\'ersic profiles. That is, the so-called 
``power-law'' galaxies not only had no distinct ``core'' like the ``core
galaxies'' do, but confusingly they do not even have power-law light profiles. 

Given that Caon et al.\ (1993) and D'Onofrio et al.\ (1994) had established
that the S\'ersic function fits the brightness profiles of elliptical galaxies
remarkably well over a large dynamic range (see the figures in Bertin et al.\
2002), it is possible to confidently identify departures from these profiles
that are diagnostic of galaxy formation.  While in this and the following
section we deal with partially-depleted cores --- also referred to as 
``missing light'' --- in 
luminous galaxies (thought to be built from dissipationless mergers), the ensuing section
addresses extra central light above the inward extrapolation of the outer
S\'ersic profile (found in galaxies that have experienced dissipation and star
formation).  

Building on the work of Caon et al.\ (1993), Graham et al.\ (2003) introduced
the core-S\'ersic model, which was applied in Trujillo et al.\ (2004).  The
model represents a modification to S\'ersic's model such that it has an inner
power-law core.  This represented a dramatic change from what had gone before.
While the Nuker team (e.g\ Kormendy et al.\ 1994; Lauer et al.\ 1995, 2005;
Faber et al.\ 1997) were combining Nuker model parameters for the core with
$R^{1/4}$ model parameters for the main galaxy, Graham et al.\ (2003), Graham
\& Guzm\'an (2003), Balcells et al.\ (2003), Graham (2004) and Trujillo et
al.\ (2004) advocated the measurement of core properties, excesses and
deficits of light, measured relative to the outer S\'ersic model.

\begin{figure*}
\includegraphics[angle=-90,scale=0.6]{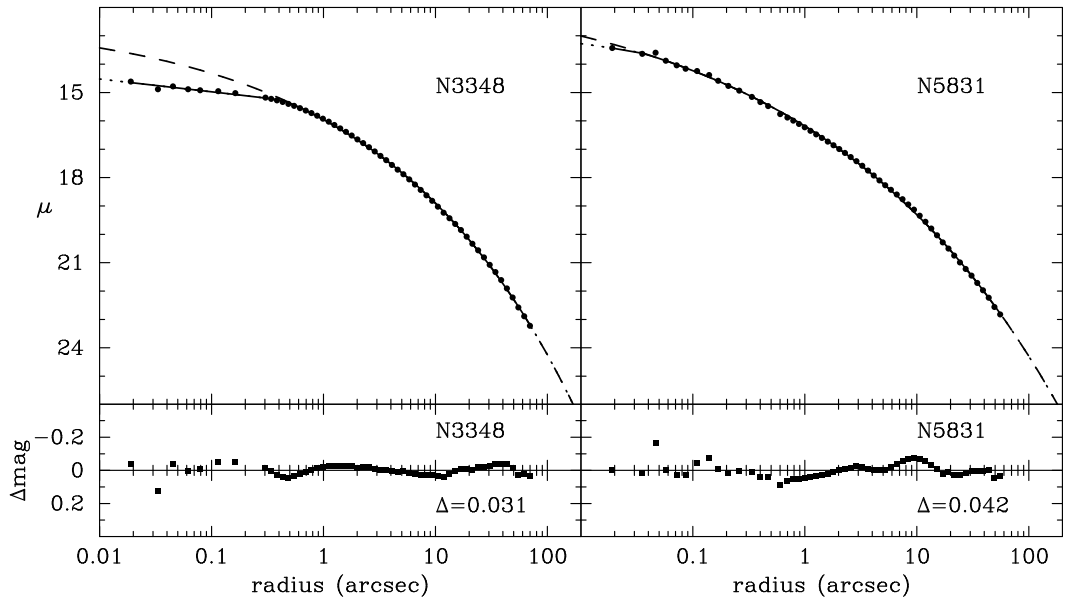}
\caption{
Left: Core-S\'ersic model (solid line) fit to the major-axis $R$-band light profile of
NGC~3348 (dots), with the dashed line showing the associated S\'ersic 
component.  The inner depleted zone 
corresponds to a stellar mass deficit of $3\times10^8 M_{\odot}$ (see Graham
2004). 
Right: NGC~5831 has, in contrast, no (obvious) partially depleted core and is
well described by the S\'ersic's model alone. The rms scatter is shown in the
lower panels. Figure taken from Graham et al.\ (2003). 
}
\label{fig-core}
\end{figure*}

For the first time, the core-S\'ersic model provided an expression capable of 
unifying the nuclear regions of galaxies with their outer regions and also 
providing stable physical quantities\footnote{The core-S\'ersic, and also 
S\'ersic, model provide robust parameters beyond the core if one 
has sufficient data to sample the curvature in the light profile, in practice 
this requires radial data out to $\sim$$1R_{\rm e}$.}. 
The model can be written as 
\begin{eqnarray}
I(R) & = & I' \left[ 1+\left( \frac{R_b}{R}\right)^{\alpha}
     \right]^{\gamma/\alpha} \nonumber \\
     &   & \times 
\exp \left\{ -b[(R^{\alpha}+R_b^{\alpha})/R_{\rm e}^{\alpha}]^{1/(\alpha
  n)}\right\},
\label{bomba}
\end{eqnarray}
where $R_b$ denotes the break--radius separating the inner power--law having
logarithmic slope $\gamma$ from the outer S\'ersic function.
The intensity $I_b$ at the break--radius is such that 
\begin{equation}
I' = I_b 2^{-(\gamma/\alpha)} \exp \left[ b(2^{1/\alpha}R_b/R_{\rm
    e})^{1/n}\right].
\label{iprime}
\end{equation}
The sixth and final parameter, $\alpha$, controls the sharpness of the
transition between the inner (power--law) and the outer (S\'ersic) regime
--- higher values of $\alpha$ indicating a sharper transition. 
Figure~\ref{fig-core} shows the core-S\'ersic model (with $\alpha=100$)
applied to NGC~3348. 

Terzi\'c \& Graham (2005) and Terzi\'c \& Sprague (2007) provide a number of 
expressions related to the potential, force and dynamics.\footnote{The
complex nature of this model has resulted in the appearance of alternate 
expressions, e.g.\ Spergel (2010, see his figure~3).}  
The core-S\'ersic model is further discussed and used by Ferrarese et al.\
(2006a,b), C\^ot\'e et al.\ (2006, 2007), Kawata, Cen \& Ho (2007) and 
Ciotti (2009). 

\subsubsection{Central Mass Deficits}\label{Sec_def}
 
The collisional construction of galaxies from the merger of lesser galaxies is
thought to be a common occurrence in the Universe.  Coupled with the presence
of a supermassive black hole (SMBH) at the heart of most galaxies (Wolfe \&
Burbidge 1970; Magorrian et al.\ 1998), 
dissipationless mergers were proposed by Begelman, Blandford \& Rees (1980;
see also Ebisuzaki, Makino, \& Okumura 1991) to explain the depleted nuclei,
i.e.\ the cores, observed in giant elliptical galaxies (e.g.\ King 1978, and
references therein).
It is thought that core-depletion is primarily due to the gravitational
slingshot (Saslaw, Valtonen, \& Aarseth 1974) effect that the coalescing SMBHs
--- from the pre-merged galaxies --- have on stars while they themselves sink
to the bottom of the potential well of the newly wed 
galaxy.\footnote{The presence of $FUV-NUV$ colour gradients in core galaxies
  such as NGC~1399 suggests that they may not have been built from major, dry
  merger events (Carter et al.\ 2011).  Additionally, the globular cluster
  specific frequency in core galaxies may be at odds with core-galaxy
  formation through equal-mass merger events because the fainter, intermediate
  luminosity elliptical galaxies may have lower specific frequencies 
  (Harris \& van den Bergh 1981). Alternative ideas for core-formation in
  giant galaxies have been proposed: Boylan-Kolchin et al.\ 2004; 
  Nipoti et al.\ 2006; Martizzi et al.\ 2011.\label{foot_dry}}

\begin{figure}
\includegraphics[angle=-90,scale=0.6]{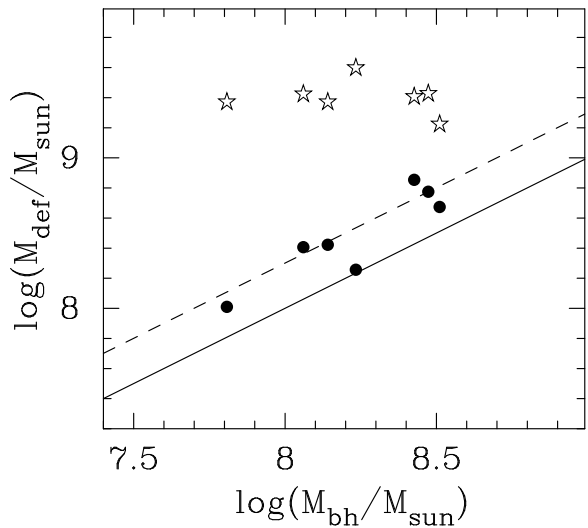}
\caption{Central mass deficit for seven core galaxies derived using the 
core-S\'ersic model (circles) and the Nuker model (stars) plotted against
each galaxy's predicted central supermassive black hole mass.  
The solid and dashed line shows $M_{\rm def}$ equal 1 and 2 $M_{\rm bh}$,
respectively.  Figure adapted from Graham (2004).} 
\label{fig_def}
\end{figure}

Theory predicts that the central mass deficit $M_{\rm def}$ should scale with
$0.5N\,M_{\rm bh}$, where $M_{\rm bh}$ is the final (merged) black hole mass
and $N$ the number of major ``dry'' (i.e.\ gas free, dissipationless) mergers
(Milosavljevi{\'c} \& Merritt 2001; Merritt \& Milosavljevi\'c 2005; Merritt
2006a,b).
Graham (2004) used the core-S\'ersic model to quantify the central deficit of
stars relative to the inward extrapolation of the outer S\'ersic profile.
Figure~\ref{fig_def} suggests that the luminous elliptical galaxies sampled
have experienced an average of 1 or 2 major dry (i.e.\ dissipationless)
mergers, a conclusion in agreement with select $\lambda$CDM models of
galaxy formation (Haehnelt \& Kauffmann 2002; Volonteri et al.\ 2003).

Quantification of the central stellar deficit relative to the inward
extrapolation of the outer S\'ersic profile has also been applied to bright
Virgo cluster galaxies by Ferrarese et al.\ (2006a) and C{\^o}t{\'e} et al.\
(2007) and with the exception of VCC 798 --- a lenticular (bulge plus disc)
galaxy --- provides similar results.
Of course when an outer disc exists, a core-S\'ersic bulge plus disc fit is
required, otherwise the disc will bias the S\'ersic
parameters.\footnote{Trujillo et al.\ (2004) and Graham (2004) 
intentionally excluded lenticular 
galaxies from their publication because it was felt that the community were
not yet ready for an 8-parameter model (6 core-S\'ersic plus 2 exponential
disc parameters). Intriguingly, VCC~798 resides near the 1-to-1 line in
Figure~\ref{fig_def} when an exponential-disc plus (core-S\'ersic)-bulge fit
is performed (Graham, unpublished work).}

\subsection{Excess nuclear light}\label{Sec_Enuc}

While galaxies brighter than $M_B\sim-20.5$ mag have partially depleted cores
(e.g.\ Faber et al.\ 1997; Rest et al.\ 2001; Graham \& Guzm\'an 2003),
fainter galaxies often contain additional stellar components at their centres.
Such excess nuclear light, above that of the underlying host galaxy, has long
been known to exist (e.g.\ Smith 1935; Reaves 1956, 1977; Romanishin, Strom
\& Strom 1977) and was systematically studied by Binggeli et al.\ (1984), van
den Bergh (1986) and Binggeli \& Cameron (1991, 1993) in a number of dwarf
elliptical galaxies.  As far back as Larson (1975) it was known that
simulations containing gas can account for these dense nuclear star clusters;
clusters which became easier to detect with the Hubble Space Telescope (e.g.\
Carollo, Stiavelli \& Mack 1998).  

\begin{figure*}
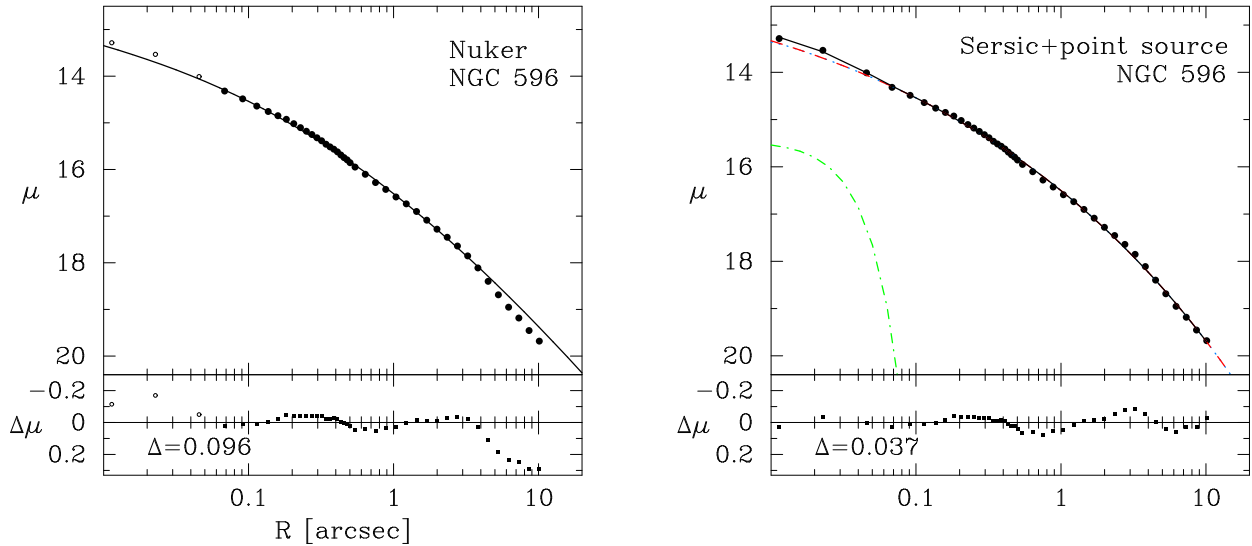

\includegraphics[angle=-90,scale=0.45]{Graham-fig5a.ps}
\hspace{1cm}
\includegraphics[angle=-90,scale=0.45]{Graham-fig5b.ps}
\caption{
Left: 5-parameter Nuker model fit to the $V$-band light-profile of the
nucleated galaxy NGC~596 after excluding the inner three data points (Lauer et al.\ 2005). 
Right: 3-parameter S\'ersic model plus 2-parameter point-source (Gaussian) 
fit to the same light profile of NGC~596.  With the same number of parameters, this 
model fits both the inner, intermediate, and outer light-profile. 
{\it Figure from Dullo \& Graham (2011, in prep.).}
}
\label{fig_596}
\end{figure*}

For many years it was common
practice to simply exclude these additional nuclear components from the
analysis of the galaxy light profile (e.g.\ Lauer et al.\ 1995; Byun et al.\
1996; Rest et al.\ 2001; Ravindranath et al.\ 2001).
Using HST data, Graham \& Guzm\'an (2003) simultaneously modelled the host
galaxy and the additional nuclear component with the combination of a S\'ersic
function for the host galaxy plus a Gaussian function for the nuclear star
cluster.  As we will see in section~\ref{Sec_Discs}, they also showed that the
lenticular galaxies in their early-type galaxy sample could be modelled via a
S\'ersic-bulge plus an exponential-disc plus a Gaussian-(star cluster)
decomposition of their light profile --- as done by Wadadekar et al.\ (1999)
with ground-based images.
Many other studies have since modeled the nuclear star clusters seen in HST
images, see Figure~\ref{fig_596}, with the combination of a nuclear component
plus a S\'ersic host galaxy (e.g.\ Grant et al.\ 2005; C\^ot\'e et al.\ 2006;
Ferrarese et al.\ 2006a; Graham \& Spitler 2009).

\begin{figure}
\includegraphics[angle=270,scale=0.3]{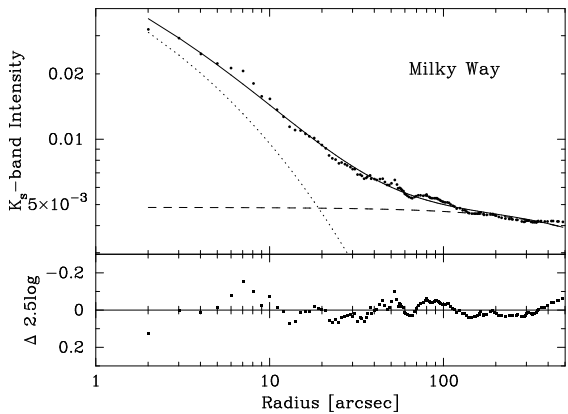}
\label{fig_milky}
\caption{
Uncalibrated, 2MASS, $K_s$-band intensity profile from the centre of the
Milky Way, taken from Sch\"odel et al.\ (2009, their Figure 2).
The nuclear star cluster is modelled with a S\'ersic function with $n=3$ 
(dotted curve) 
and the underlying host bulge with an exponential function (Kent et al.\ 1991)
that has an effective half-light radius of 
$\sim$4.5 degrees (e.g.\ Graham \& Driver 2007) and is therefore basically a
horizontal line. One parsec equals 25 arcseconds. 
{\it Figure from Graham \& Spitler (2009).} 
}
\end{figure}

While Graham \& Guzm\'an (2003) and C\^ot\'e et al.\ (2006) found that some
nuclear star clusters could actually be resolved, it is not yet established
what mathematical function best describes their radial distribution of stars.  
The closest example we have to study is of course the 30 
million solar mass nuclear star cluster at the centre of the Milky Way
(Launhardt et al.\ 2002).  Graham \& Spitler (2009) provided what may have
been the first ever analysis after allowing for the significant contamination
from bulge stars.  Although they found that the cluster could be well
described by a S\'ersic index with $n=3$ (see Figure~\ref{fig_milky}, it
remains to be tested how well a King model can describe the data, or at least
the old stellar population known to have a core (e.g.\ Genzel et al.\ 1996). 
The excess nuclear light in M32 has also been well fit with an $n=2.3$ S\'ersic function
(Graham \& Spitler) but this too may yet be better described by a King model.
What is apparent is that neither the underlying bulge nor the nuclear
component are described by a power-law.  Theories which form or assume 
such power-law cusps appear to be at odds with current observations.

\subsection{Excess halo light}\label{Sec_cD}

This section shall be somewhat cursory given R.Bower's detailed chapter on
clusters and the intracluster medium in this volume.

Brightest cluster galaxies (BCGs), residing close to or at the centres of
large galaxy clusters, have long been recognised as different from less
luminous elliptical galaxies: their light profiles appear to have excess flux
at large radii (e.g.\ Oemler 1976; Carter 1977; van den Bergh 1977; Lugger
1984; Schombert et al.\ 1986).
However, before exploring this phenomenon, it is important to recall that 
the light profiles of large galaxies have S\'ersic indices $n$ greater than 4
(e.g.\ Caon et al.\ 1993; Graham et al.\ 1996).\footnote{As $n \rightarrow \infty$,
the S\'ersic model can be approximated by a power-law (see Graham \& Driver
2005).}. Subsequently, at large radii in big elliptical galaxies, 
there will be excess flux above that of 
an $R^{1/4}$ model fit to some limited inner radial range. 

An initially puzzling result from the Sloan Digital Sky Survey ({\sl SDSS};
York et al.\ 2000) survey was the lack of light profiles with S\'ersic $n >
5$--6 (Blanton et al.\ 2005a).  This was however soon resolved when Blanton et
al.\ (2005b), Mandelbaum et al.\ (2005) 
and Lisker (2005, 2006b, 2007) pointed 
out a serious sky-subtraction problem with the early SDSS Photometric Pipeline
(photo: Ivezi\'c et al.\ 2004).
The sky-value to be subtracted from each galaxy had been measured to close to the
galaxy in question, and because galaxies with large S\'ersic indices possess
rather extended light-profiles, their outer galaxy light was actually measured
and then subtracted from these galaxies.  As a result, the high S\'ersic index
light-profiles were erroneously erased and missed from the SDSS.  This
resulted in the $R^{1/4}$ model appearing to provide good fits to bright 
elliptical galaxies, and consequently a series of papers based on structurally
biased radii, magnitudes and surface brightnesses.

Bearing in mind that large elliptical galaxies have high S\'ersic indices, it
is important to distinguish between (i) an inadequacy of the $R^{1/4}$ model
to describe what is actually a single $R^{1/n}$ profile, and (ii) a distinct
physical component such as an envelope of diffuse halo light surrounding a
central galaxy.
Early quantitative photometry of cD galaxies (supergiant D galaxies: e.g.\ 
Matthews et al.\ 1964; Morgan \& Lesh 1965) 
suggested the presence of an inner $R^{1/4}$ spheroid plus an outer
exponential corona (de Vaucouleurs 1969; de Vaucouleurs \& de Vaucouleurs
1970).  One should however question if this outer corona is a distinct entity
or not. To answer this in the affirmative, astronomers can point to how the
light profiles can display inflections marking the transition from BCG light
to intracluster light.
Gonzalez et al.\ (2005) additionally showed that the inflection in the light
profiles of many BCGs was also associated with an a change in the ellipticity
profile, signalling the switch from BCG light to intracluster light.

Gonzalez et al.\ (2005) and Zibetti et al.\ (2005) chose to model their BCG
sample using an $R^{1/4} + R^{1/4}$ model to describe the inner galaxy plus
the outer halo light.  However, given that elliptical galaxies are better
described by the $R^{1/n}$ model, and the desire to measure the actual
concentration of halos rather than assign a fixed $R^{1/4}$ profile, Seigar et
al.\ (2007) fitted an $R^{1/n} + R^{1/n}$ model to the light profiles of five
BCGs.  Not surprisingly, they found that the $R^{1/4} + R^{1/4}$ model was not
the optimal representation.  An $R^{1/n}$-galaxy plus an exponential-halo
model was found to provide the optimal fit in three instances, with an
additional galaxy having no halo detected.  The associated galaxy-to-halo luminosity
ratios can be found there.  This re-revelation of an exponential model, rather
than an $R^{1/4}$ model, describing the halo has since been confirmed by
Pierini et al.\ (2008).  Intriguing is that the halo does not trace the
NFW-like dark-matter halo density profiles produced in $\Lambda$CDM
simulations (section~\ref{Sec_dark}).  Stellar halos around non-BCG
galaxies have also now been reported to display an exponential radial distribution
(e.g.\ Gadotti 2011; Tal \& van Dokkum 2011).

\section{Structure related scaling relations}\label{Sec_E} 

While it is common practice, and somewhat helpful, to call elliptical galaxies
fainter than about $M_B = -18$ mag by the term ``dwarf elliptical'' (Sandage
\& Binggeli 1984), it should be noted, and this section will reveal, that on
all measures they appear to be the low-mass end of a continuous sequence which
unifies dwarf and normal/luminous elliptical galaxies.  Not only is this true
in terms of their structural properties (e.g.\ Binggeli et al.\ 1984; Graham
\& Guzm\'an 2003; Gavazzi et al.\ 2005; Ferrarese et al.\ 2006a; C\^ot\'e et
al.\ 2006, 2007, 2008; Misgeld et al.\ 2008, 2009; Janz \& Lisker 2009; Graham
2010; Chen et al.\ 2010; Glass et al.\ 2011) but even the degree of
kinematically distinct components is similar (Chilingarian 2009).

There are many important relations between stellar luminosity, colour,
metallicity, age, and dynamics that reveal a continuous and
linear behaviour uniting dwarf and giant elliptical galaxies (e.g.\ Caldwell
1983; Davies et al.\ 1983; Binggeli et al.\ 1984; Bothun et al.\ 1986; Geha et
al.\ 2003; Lisker \& Han 2008). 
However, Kormendy et al.\ (2009, their section~8) dismiss all of these apparently
unifying relations by claiming that they must not be sensitive to different
physical processes which they believe have produced a dichotomy between dwarf
and ordinary elliptical galaxies.  They claim that it is only relations which
show an apparent different behavior at the faint and bright end that are
sensitive to the formation physics and the remainder are not relevant.  This
section explains why such non-linear relations are actually a consequence of
the (dismissed) linear relations, and as such these non-linear relations actually support a
continuum between dwarf and ordinary elliptical galaxies. 

To begin, it should be reiterated that (dwarf and ordinary) elliptical
galaxies --- and the bulges of disc galaxies (section~\ref{Sec_BD}) --- do not have
structural homology. Instead, they have a continuous range of stellar
concentrations --- quantified by the S\'ersic index $n$ (Davies et al.\ 1988;
Caon et al.\ 1993; 
D'Onofrio et al.\ 1994; Young \& Currie 1994, 1995; Andredakis et al.\ 1995)
--- that varies linearly with both stellar mass and central surface brightness
(after correcting for central deficits or excess light).  A frequently
unappreciated consequence of these two linear relations is that relations
involving either the effective half-light radius ($R_{\rm e}$) or the
effective surface brightness ($\mu_{\rm e}$), or the mean surface brightness
within $R_{\rm e}$ ($\langle \mu \rangle_{\rm e}$), will be non-linear.  Such
curved relations have often been heralded as evidence that two different
physical processes must be operating
because the relation is not linear and has a different slope at either end.
To further complicate matters, sample selection which includes faint and
bright elliptical galaxies, but excludes the intermediate-luminosity
population, can effectively break such continuously curved relations into two
apparently disconnected relations, as can selective colour-coding.

There are three distinct types of (two-parameter) relations involving the properties of
elliptical galaxies: (i) linear relations which are taken to reveal the unified nature 
of dEs and Es; (ii) curved relations revealing a continuity that had in the 
past been mis-interpreted to indicate that distinct formation
process must be operating; and (iii) broken relations which imply that two physical mechanisms are operating. 
In the following sections we shall learn how the linear relations 
result in the existence of curved relations when using effective radii and
surface brightnesses.  We shall also see that the transition in the broken 
relations occurs at $M_B \approx -20.5$ mag and thus has nothing to do with
the previously held belief that dEs and Es are two distinct species separated
at $M_B = -18$ mag (Wirth \& Gallager 1984; Kormendy 1985b, 2009).

\subsection{Linear relations}\label{Sec_Line}

This section introduces two key relations, from which a third can be derived, 
involving structural parameters.  They are the 
luminosity-concentration ($L$--$n$) relation and the luminosity-(central
density)\footnote{Here the ``central density'' refers to the density prior to
core depletion in giant elliptical galaxies or the growth of additional
nuclear components in smaller elliptical galaxies.} ($L$--$\mu_0$) relation.
We shall use the S\'ersic shape parameter $n$ to quantify the central
concentration of the radial light profile, and use the projected central
surface brightness $\mu_0$ as a measure of the central density.\footnote{To
convert from the surface density to the internal density, one can use
equation~4 from Terzi\'c \& Graham (2005).}  

It is noted that one can expect a certain level of scatter in the $L$--$n$ and
$L$--$\mu_0$ diagrams because both the central density and the radial
concentration of stars that one observes depends upon the chance orientation
of one's triaxial galaxy (Binney 1978).
This is of course also true for measurements of effective surface brightness, 
half-light radii, velocity dispersions, etc. 
To have the best relations, it is important that we use S\'ersic parameters
from elliptical galaxies rather than parameters from single-S\'ersic fits to
samples of elliptical and lenticular galaxies given the two-component (2D-disc
plus 3D-bulge) nature of the lenticular galaxies.
Given the offset nature of bulges and elliptical galaxies in the $L$--$n$
diagram (e.g.\ Graham 2001; see also M\"ollenhoff \& Heidt 2001) it is also 
important that bulges not be combined in this section's analysis of elliptical 
galaxy scaling relations.

\subsubsection{Luminosity-(central surface brightness) relation}\label{Sec_M-mu0}

\begin{figure*}
\begin{center}
\includegraphics[angle=270,scale=0.68]{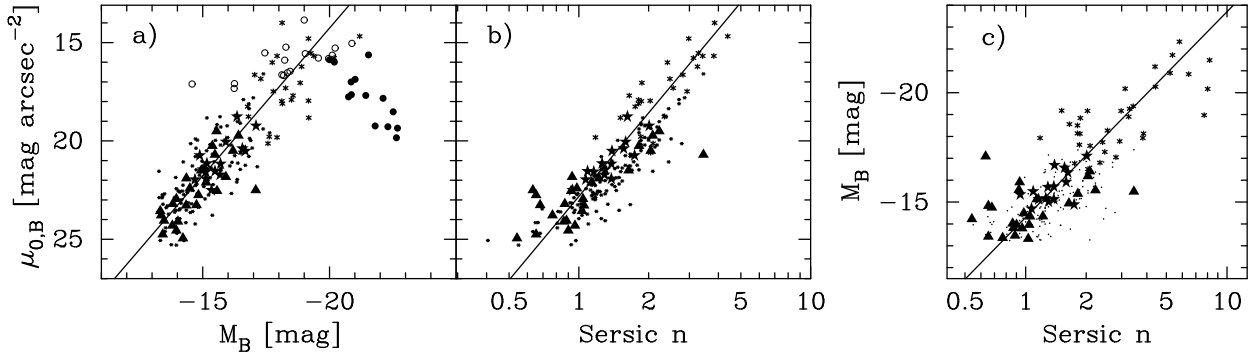}
\caption{Linear relations between the observed B-band central surface
  brightness ($\mu_{0,B}$) with a) the absolute B-band magnitude $M_B$ and b)
  the S\'ersic index $n$ for a sample of elliptical galaxies.  Panel c) shows
  the galaxy magnitudes versus the S\'ersic indices, with the line given by
  equation~\ref{Eq_M-n}.  The ``core galaxies'' (large filled circles) with
  partially depleted cores can be seen to have lower central surface
  brightnesses than the relation in panel a).  Inward extrapolation of their
  outer S\'ersic profile yields $\mu_0$ values which follow the linear
  relation, as previously noted by Jerjen \& Binggeli (1997).
  The data have come from the compilation by Graham \& Guzm\'an (2003, their
  figure~9).  Dots represent dE galaxies from Binggeli \& Jerjen (1998),
  triangles represent dE galaxies from Stiavelli et al.\ (2001), large stars
  represent Graham \& Guzm\'an's (2003) Coma dE galaxies, asterix represent
  intermediate to bright E galaxies from Caon et al.\ (1993) and D'Onofrio et
  al.\ (1994), open circles represent the so-called ``power-law'' E galaxies
  from Faber et al.\ (1997) and the filled circles represent the ``core'' E
  galaxies from these same Authors.  }
\label{Fig_abc}
\end{center}
\end{figure*}

Caldwell (1983; his Figure 6) and Bothun et al.\ (1986, their figure~7)
revealed that, fainter than $M_B \approx -20.5$ mag, there is a continuous
linear relation between luminosity and central surface brightness.
Furthermore, Binggeli et al.\ (1984, their figure~11) 
and Binggeli \& Cameron (1991, their figure~9 and 18) 
revealed that, when using the inward extrapolation of King models, 
the $L$--$\mu_0$ relation is continuous and roughly linear from 
$-12 > M_B > -23$ mag.  This same general result was also highlighted by 
Jerjen \& Binggeli (1997) and 
Graham \& Guzm\'an (2003) when using the inward extrapolation of the S\'ersic model.
The benefit of this approach is that one's central surface brightness 
is not biased by the presence of a depleted 
core or any additional nuclear components within the host galaxy. 
Figure~\ref{Fig_abc}a displays the elliptical galaxy ($M_B$, $\mu_0$) data set 
from Graham \& Guzm\'an (2003) fit by the expression
\begin{equation}
M_B = 0.67\mu_{0,B} - 29.5.
\label{Eq_M-mu0}
\end{equation}

The actual central surface brightness of the luminous ``core galaxies'' is
shown in Figure~\ref{Fig_abc}a, 
rather than the value obtained from the inward extrapolation of their outer
S\'ersic profile.  As such these ``core galaxies'' were excluded from the fit,
but see the discussion in section~\ref{Sec_bent}. 
As an aside, if the central supermassive black hole mass $M_{\rm bh}$ in
elliptical galaxies is directly related to the central stellar density (see
Graham \& Driver 2007), then the connections between $M_{\rm bh}$ and the
global galaxy properties, such as total mass and velocity dispersion, may be
secondary.

\subsubsection{Luminosity-concentration relation}\label{Sec_M-n}

The linear relation between luminosity and S\'ersic index, or strictly
speaking the logarithm of these quantities, has been shown many times (e.g.\
Young \& Currie 1994; Jerjen \& Binggeli 1997; Graham \& Guzm\'an 2003; 
Ferrarese et al.\ 2006a).  This continuous relation between magnitude and
concentration\footnote{Graham et al.\ (2001) contains a review of various
  concentration indices used over the decades, while Trujillo et al.\ (2001)
  was the first to quantify the monotonic relation between S\'ersic index and concentration.}
for elliptical galaxies had of course been recognised before (e.g.\
Ichikawa et al.\ 1986, their figure~11). 
The following $M_B$--$n$ expression is shown in 
Figure~\ref{Fig_abc}c, again matched to the sample of elliptical galaxies compiled
by Graham \& Guzm\'an (2003). 
\begin{equation}
M_B = -9.4\log(n) - 14.3.
\label{Eq_M-n}
\end{equation}
Graham \& Guzm\'an (2003) excluded two-component 
lenticular galaxies fit by others with a single-component S\'ersic model. 
It may be prudent to continue to exclude these galaxies even after a 
S\'ersic-bulge plus exponential disc fit because the $M_B$--$n$ relation
defined by bulges, at least in spiral galaxies, is different to that 
defined by elliptical galaxies (Graham 2001, his figure~14).

\subsubsection{Concentration-(central surface brightness) relation}

Combining the above two equations provides an expression between central
surface brightness and S\'ersic index such that 
\begin{equation}
\mu_0 = 22.8 - 14.1 \log(n), 
\label{Eq_mu0-n}
\end{equation}
which is shown in Figure~\ref{Fig_abc}b, where it can be seen to be 
roughly applicable for values of $n \gtrsim 1$.

\subsection{Curved relations}\label{sec_curved}

This section explains why, as a direct result of the above linear
relations --- which unite dwarf and giant elliptical galaxies --- expressions
involving either the effective half-light radius $R_{\rm e}$, the associated
effective surface brightness $\mu_{\rm e}$ at this radius, or the mean surface
brightness $\langle \mu \rangle _{\rm e}$ enclosed within this radius, are 
curved.  

\subsubsection{Luminosity-(effective surface brightness) relation}

The following analysis is from Graham \& Guzm\'an (2003). 

Given the empirical $M_B$--$n$ relation (equation~\ref{Eq_M-n}), one knows what the
expected value of $n$ is for some value of $M_B$.  One can then convert the
empirical $M_B$--$\mu_0$ relation (equation~\ref{Eq_M-mu0}) into an 
$M_B$--$\mu_{\rm e}$ relation using the exact relation
between $\mu_0$ and $\mu_{\rm e}$ which depends only on the
value of $n$ (equation~\ref{Eq_mue-mu0}).  Doing so, 
one obtains the expression
\begin{eqnarray}
\mu_{\rm e} & = & 1.5M_B + 44.25 + 1.086b,  \nonumber \\
            & = & 1.5[M_B+14.3] + 22.8 + 1.086b,
\label{Eq_M-Mue}
\end{eqnarray}
where $b \approx 1.9992n-0.3271$ and 
equation~\ref{Eq_M-n} is used to replace $n$ in terms of $M_B$, such
that $n = 10^{ -(14.3+M_B)/9.4}$.

\begin{figure*}
\begin{center}
\includegraphics[angle=270,scale=0.7]{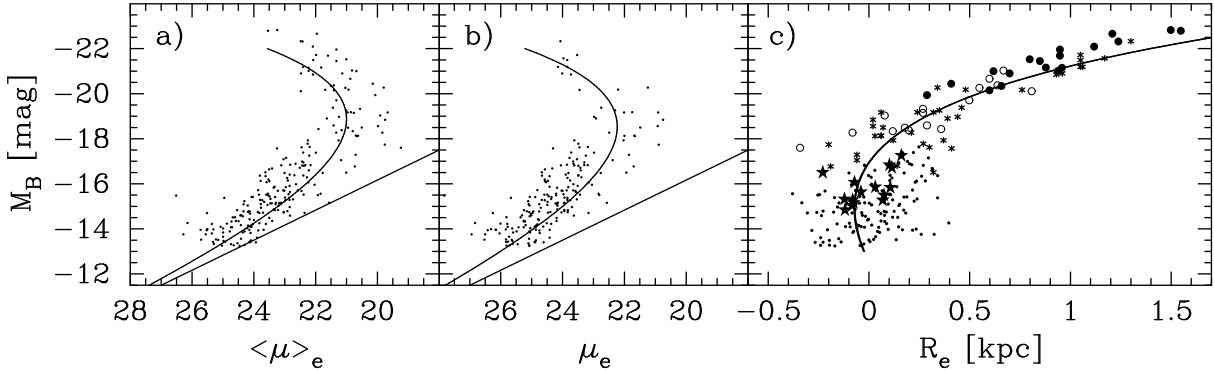}
\caption{
Elliptical galaxy B-band magnitude versus a) mean effective surface brightness
(equation~\ref{Eq_M-Muav}), b) effective surface brightness (equation~\ref{Eq_M-Mue}),
and c) effective radius (equation~\ref{Eq_M-Re}).
The continuous, curved relations are predictions from the observed linear relations in
Figure~\ref{Fig_abc}.  The slight mismatch seen here arises from the rough
fitting-by-eye of the linear relations. 
The data points have come from the compilation by Graham \& Guzm\'an (2003, their
figure~9). 
}
\label{Fig_Magic}
\end{center}
\end{figure*}

One can similarly convert the 
empirical $M_B$--$\mu_0$ relation (equation~\ref{Eq_M-mu0}) into an 
$M_B$--$\langle \mu \rangle _{\rm e}$ relation using the exact relation
between $\mu_0$ and $\langle \mu \rangle _{\rm e}$ which also depends only on the
value of $n$ (equation~\ref{Eq_mue-mue}).  Doing this, one obtains the
expression
\begin{equation}
\langle \mu \rangle _{\rm e} = 1.5M_B + 44.25 + 1.086b -2.5\log \left[
  \frac{{\rm e}^{b}n\Gamma(2n)}{b^{2n}} \right], 
\label{Eq_M-Muav}
\end{equation}
where again $b \approx 1.9992n-0.3271$ and equation~\ref{Eq_M-n} is used to
replace $n$ in terms of $M_B$. 
These curves can be seen in Figure~\ref{Fig_Magic} (adapted from Graham
\& Guzm\'an 2004). 

Binggeli et al.\ (1984, their figure~8) and Capaccioli \& Caon (1991) 
previously showed with empirical data that the $M_B$--$\langle \mu 
\rangle _{\rm e}$ relation is curved. 
What was new in Graham \& Guzm\'an (2003) was the explanation. 
In the past, evidence of non-linear relations involving parameters from dwarf
and giant elliptical galaxies were heralded as evidence of a dichotomy.
Coupled with galaxy sample selection that excluded the intermediate
population, and therefore resulted in two apparently disconnected relations, 
acted to further convince some that they were dealing with two classes of
object.\footnote{A simple one-dimensional example of sample bias would be a survey of
the average physical properties, such as size or mass, of people at a primary
school. One would measure the properties of children and adults, but miss the
bridging population which reveals a continuity and thus unification of the
species.}

\subsubsection{Size-Luminosity relation}

Now that we know how to play this game, one can additionally make predictions
for relations involving the effective radius $R_{\rm e}$ because we know that
the luminosity $L=2\pi \langle I \rangle _{\rm e} R_{\rm e}^2$, with $\langle
\mu \rangle _{\rm e} = -2.5\log \langle I \rangle _{\rm e}$.  As explained in
Graham \& Worley (2008, their section~5.3.1), one can derive the
size-luminosity relation such that 
\begin{equation}
\log R_e [{\rm kpc}] =  
\frac{M_B}{10} + 1.066 + 0.434n + 0.5\log \left[
  \frac{b^{2n}} {{\rm e}^{b}n\Gamma (2n)} \right]
\label{Eq_M-Re}
\end{equation}
for $0.5 < n < 10$, with $b \approx 1.9992n-0.3271$ 
and where equation~\ref{Eq_M-n} is used to replace $n$ in terms of $M_B$.
This size-luminosity relation for elliptical galaxies is shown in 
Figure~\ref{Fig_Magic}c along with real galaxy data. 

Binggeli et al.\ (1984, their figure~7; cf.\ Misgeld \& Hilker 2011, their
figure~1) also demonstrated, with empirical data, that the $L-R_e$ relation for
dwarf and giant elliptical galaxies is curved.  Their diagram, in addition to
Figure~\ref{Fig_Magic}c seen here, reveals why studies which only sample
bright elliptical galaxies are often contempt to simply fit a straight line
(e.g.\ Kormendy 1977b).  The explanation for why this 
happens is of course akin to approximating the Earth as flat when one
(forgivably) does not sample enough of what is actually a curved profile.  As
Graham et al.\ (2006, their figure~1b) re-revealed recently, and as reiterated by
Bernardi et al.\ (2007), a sample of massive elliptical galaxies will have a
steeper size-luminosity relation than a sample of ordinary elliptical
galaxies, which will in turn have a steeper size-luminosity relation than a
sample of dwarf elliptical galaxies because the size-luminosity relation is
curved.
Graham \& Worley (2008) explains why the $L-R_e$ relation given by
equation~\ref{Eq_M-Re}, based on two linear relations and the functional form
of S\'ersic's model, is curved.

Interestingly, due to the linear relations between magnitude, central surface
brightness and the logarithm of the S\'ersic exponent $n$, the use of faint isophotal radii results in
what is roughly a linear size-luminosity relation (e.g.\ Oemler 1976; Strom \&
Strom 1978; Forbes et al.\ 2008; van den Bergh 2008; Nair et al.\ 2011), with the 
bright-end slope dependent on the adopted isophotal limit or Petrosian radius
used.  The implications of this important observation 
shall be detailed elsewhere. 

Helping to propagate the belief that dwarf and ordinary elliptical galaxies
are distinct species, Dabringhausen et al.\ (2008) and Lisker (2009) 
fit a double power-law to their curved size-luminosity relation for dwarf and
ordinary elliptical galaxies, thus yielding distinct slopes at the faint
and bright end. 
In addition, the interesting study by Janz \& Lisker (2008) 
reported small deviations from the predicted curved relation.  However, 
galaxies that are well described by S\'ersic's function and which follow
linear $M$--$n$ and $M$--$\mu_0$ relations {\it must} follow a single curved
$M$--$R_{\rm e}$ relation.  The deviations that they found are therefore
mirroring a) the inadequacy of the fitted linear relations to the $M$--$n$ and
$M$--$\mu_0$ distribution (a point noted in the caption of Figure~\ref{Fig_Magic}) and/or b)
poor fitting S\'ersic models to their sample of elliptical {\it and} disc
galaxies.  Adding uncertainties to the linear relations in
section~\ref{Sec_Line}, and propagating those through to the predicted
$M$--$R_{\rm e}$ relation is required before we can claim evidence of
significant deviations.  However, searching for such second order effects may
indeed be interesting given the different types of dwarf galaxies that are
emerging (Lisker et al.\ 2007). 

\subsubsection{Size-concentration relation}

\begin{figure}
\includegraphics[angle=270,scale=0.43]{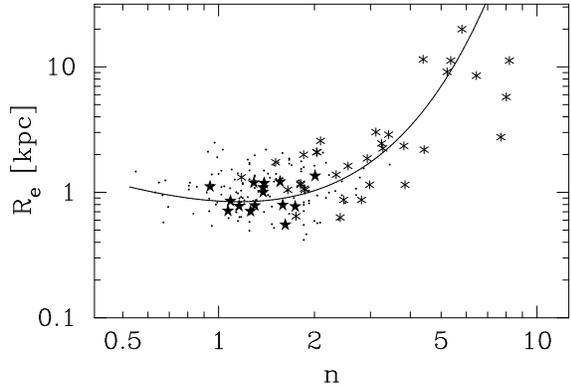}
\caption{Equation~\ref{Eq_n-Re} is over-plotted empirical data.
Symbols have the same meaning as in Figure~\ref{Fig_abc}. 
}
\label{Fig_n_R}
\end{figure}

One can additionally derive an expression relating $R_{\rm e}$ and $n$ by 
substituting the magnitude from the empirical 
$M_B$--$n$ relation, expressed in terms of $n$ (equation~\ref{Eq_M-n}), into 
the size-luminosity relation (equation~\ref{Eq_M-Re}) to give 
\begin{eqnarray}
\log R_e [{\rm kpc}] & = &
0.434n - 0.364 -0.94\log(n)  \nonumber \\
 & & + 0.5\log \left[ \frac{b^{2n}} {{\rm e}^{b}n\Gamma (2n)} \right]. 
\label{Eq_n-Re}
\end{eqnarray}
While the $M_B$--$n$ relation is linear, 
the $R_{\rm e}$--$n$ relation is curved, as can be seen in
Figure~\ref{Fig_n_R}. 

In passing it is noted that the form of this relation (equation~\ref{Eq_n-Re})
matches the bulge data from Fisher \& Drory (2010, their Figure~13).  They
interpret the departure of the low-$n$ bulges ($n<2$) from the approximately
linear relation defined by the high-$n$ bulges ($n>2$) to indicate that a
different formation process is operating to produce the less concentrated
``pseudobulges''.  
However, 
based upon linear unifying relations that span the artificial $n=2$ divide,
we know that this $R_{\rm e}$--$n$ relation must be curved. 
Without an understanding of this relation, and other curved relations
(e.g. Greene, Ho \& Barth 2008), they
have at times been misinterpretted and used to claim the existence of different
physical processes (see section~\ref{sec_pseudo} for a discussion of
pseudobulges: Hohl 1975 and references therein). 

It may be worth better defining the behavior of the $R_{\rm e}$--$n$ relation
at small sizes in Figure~\ref{Fig_n_R}.  The data from Davies et al.\ (1988)
suggests than when $n=0.5$, values of $R_{\rm e}$ may range from 1 kpc down to
0.2 kpc (Caon et al.\ 1993, their figure~5).  Such a reduction to the
flattening of the $R_{\rm e}$--$n$ distribution, below $n\approx 1$, may in
part arise from the inclusion of dwarf spheroidal galaxies (see Misgeld \&
Hilker 2011, their figure~1).

\subsubsection{Size-(effective surface brightness) relation}\label{Sec_Korm}

\begin{figure}
\begin{center}
\includegraphics[angle=270,scale=0.51]{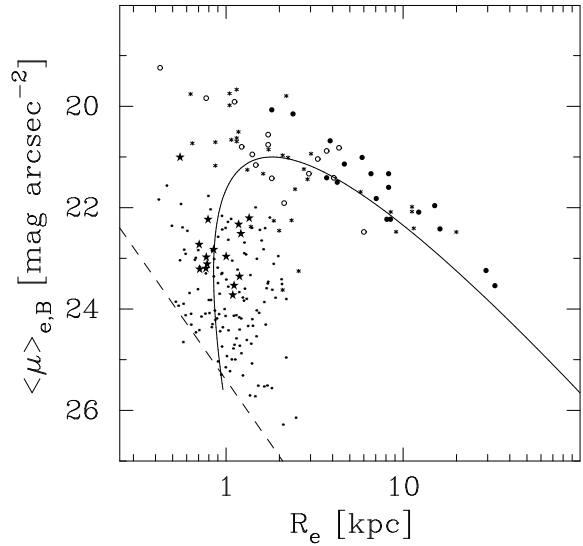}
\caption{Due to the observed linear relations in Figure~\ref{Fig_abc}, 
  the relation between the effective 
  radius ($R_{\rm e}$) and the mean surface brightness within this radius
  ($\langle \mu \rangle_{\rm e}$) is highly curved for elliptical galaxies.  
  The dashed line shows the $M_B =-13$ mag limit for the Virgo cluster.
}
\label{Fig_Korm}
\end{center}
\end{figure}

As discussed in Graham (2010), 
the first two linear relations in Figure~\ref{Fig_abc} 
naturally explain the curved $\langle \mu \rangle _{\rm e}$--$R_{\rm e}$ 
relation in Figure~\ref{Fig_Korm}. 
From the empirical $\mu_0$--$n$ relation (equation~\ref{Eq_mu0-n}, 
Figure~\ref{Fig_abc}b), one can convert $\mu_0$ into 
$\langle \mu \rangle_{\rm e}$ using equations~\ref{Eq_mue-mu0} and 
\ref{Eq_mue-mue}. 
The effective radius $R_{\rm e}$ is acquired by matching the empirical 
$M_B$--$\mu_0$ relation (equation~\ref{Eq_M-mu0}, 
Figure~\ref{Fig_abc}a) with the absolute magnitude formula 
\begin{equation}
M = \langle \mu \rangle _{\rm e} - 2.5\log(2\pi R_{\rm e,kpc}^2) - 36.57,
\label{Eq_mag_form}
\end{equation} 
(see Graham \& Driver 2005, their equation~12).  Eliminating the absolute
magnitude gives the expression 
\begin{equation}
\log R_{\rm e} = \frac{1}{5}\left\{ \frac{\langle \mu \rangle_{\rm e}}{3} -9.07
+0.72b -1.67\log\left(\frac{n{\rm e}^b\Gamma(2n)}{b^{2n}}\right) \right\}, 
\label{Eq_MuR}
\end{equation}
in which we already know the value of $n$ associated with each value of 
$\langle \mu \rangle_{\rm e}$. 
This is achieved by (again) using the empirical $\mu_0$--$n$ relation
(equation~\ref{Eq_mu0-n}) with equations~\ref{Eq_mue-mu0} and
\ref{Eq_mue-mue}, such that 
\begin{equation}
\langle \mu \rangle_{\rm e} = 22.8 + 1.086b -14.1\log(n) -2.5\log \left[
  \frac{n{\rm e}^{b}\Gamma(2n)}{b^{2n}} \right]
\label{Eq_mu-n}
\end{equation}
and $b \approx 1.9992n-0.3271$. 
Equation~\ref{Eq_MuR}, obtained from two linear relations involving S\'ersic
parameters, is a curved relation that is shown in Figure~\ref{Fig_Korm}.
Overplotted this predicted relation are data points from real galaxies. 

For those who may have the S\'ersic parameter set ($R_{\rm e}, \mu_{\rm e}, n$), one
can use equation~\ref{Eq_mue-mue} to convert $\mu_{\rm e}$ into $\langle \mu
\rangle_{\rm e}$ if one wishes to compare with the relation given by
equation~\ref{Eq_MuR}.  For those who may have the parameter set ($R_{\rm e},
\mu_{\rm e}$), perhaps obtained with no recourse to the S\'ersic
model, equation~\ref{Eq_MuR} can easily be adjusted using
equation~\ref{Eq_mue-mue} to give a relation between $R_{\rm e}$ and 
$\mu_{\rm e}$ such that
\begin{equation}
\log R_{\rm e} = \frac{\mu_{\rm e}}{15}-1.81
  -0.5\log\left(\frac{n{\rm e}^b\Gamma(2n)}{10^{0.29b}\, b^{2n}}\right), 
\label{Eq_MueR}
\end{equation}
where the value of $n$ associated with the value of $\mu_{\rm e}$ is given by
\begin{equation}
\mu_{\rm e} = 22.8 + 1.086b -14.1\log(n). 
\label{Eq_muon}
\end{equation}

To summarise, 
due to the linear relations in Figure~\ref{Fig_abc} which connect dwarf and
ordinary elliptical galaxies across the alleged divide at $M_B = -18$ mag
(Kormendy 1985), or at $n=2$ (Kormendy \& Kennicutt 2004), 
coupled with the smoothly varying change in light profile shape as a function
of absolute magnitude, the $\langle \mu \rangle_{\rm e}$-$R_{\rm e}$ and $\mu_{\rm
e}$-$R_{\rm e}$ relations are expected to be curved (Figure~\ref{Fig_Korm}), as previously 
shown with empirical data by, for example, Capaccioli \& Caon (1991). 
This also explains why the fitting of a linear relation to 
($R_{\rm e}, \mu_{\rm e}$) data by Hoessel et al.\ 
(1987) resulted in slopes that depended on their galaxy sample magnitude.  

The Kormendy relation is a tangent to the bright arm of what is actually a
curved distribution defined by the relation given by equation~\ref{Eq_MuR}
that is taken from Graham (2010). The apparent deviant nature of the dwarf elliptical
galaxies from the approximately linear section of the bright-end of the
$\langle \mu \rangle_{\rm e}$-$R_{\rm e}$ distribution does not necessitate
that a different physical process be operating.
Moreover, as noted by Graham \& Guzm\'an (2004) and Graham (2005), galaxies
which appear to branch off at the faint end of the Fundamental Plane
(Djorgovski \& Davis 1987) --- the flat portion at the bright end of a curved
hypersurface --- also need not have formed from different physical mechanisms.
Simulations that assume or reproduce a linear $\langle \mu \rangle_{\rm
e}$-$R_{\rm e}$ or $\mu_{\rm e}$-$R_{\rm e}$ relation, across too great a
magnitude range, have failed to mimic the continuous curved distribution
defined by real elliptical galaxies.  The same remark is true for simulations
of the `Fundamental Plane'.

\subsection{Broken relations}\label{Sec_bent}

\subsubsection{Luminosity-(central surface brightness) relation}\label{L-Mu0}

While the relation between a galaxy's absolute magnitude and {\it extrapolated}
central surface brightness is remarkably linear (section~\ref{Sec_M-mu0}),
there is a clear break in this relation 
when using the {\it actual} central surface brightness at the luminous end of the
distribution 
(Figure~\ref{Fig_abc}a).  This departure from the $M_B$-$\mu_0$
relation by elliptical galaxies brighter than $M_B \approx -20.5$ mag ($M >
0.5$-$1\times 10^{11} M_{\odot}$) was addressed by Graham \& Guzm\'an (2003) in
terms of partially depleted cores relative to the outer S\'ersic profile (see
also Graham 2004; Trujillo et al.\ 2004; Merritt \& Milosavljevi\'c 2005; Ferrarese
et al.\ 2006a; C\^ot\'e et al.\ 2007).  This transition has nothing to do with
the alleged divide between dwarf and giant elliptical galaxies at around $M_B
= -18$ mag, but is instead directly related with the S\'ersic versus core-S\'ersic
transition at around $M_B = -20.5$ mag. 

As noted in section~\ref{Sec_def}, 
such partially depleted cores in luminous core-S\'ersic galaxies are thought 
to have formed from dry, dissipationless galaxy merger events involving the central coalescence
of supermassive black holes (Begelman, Blandford, \& Rees 1980; Ebisuzaki,
Makino, \& Okumura 1991; but see footnote~\ref{foot_dry}) 
and resulted in Trujillo et al.\ (2004) advocating a
``new elliptical galaxy paradigm'' based on the presence of a central stellar
deficit versus either none or an excess of light, an approach embraced by
Ferrarese et al.\ (2006a), C\^ot\'e et al.\ (2007) and others. 

Further evidence for a division at $M_B = -20.5$ mag comes from the tendency
for the brighter galaxies to be anisotropic, pressure supported elliptical
galaxies having boxy isophotes, while the less luminous early-type galaxies
may have discy isophotes and often contain a rotating disc (e.g.\ Carter 1978,
1987; Davies et al.\ 1983; Bender et al.\ 1988; Peletier et al.\ 1990; Jaffe
et al.\ 1994).  Core galaxies also tend to be more radio loud and have a
greater soft X-ray flux (e.g.\ Ellis \& O'Sullivan 2006; Pellegrini 2010; 
Richings, Uttley \& K\"ording 2011, and references therein).

It was, in part, from a diagram of central surface brightness versus magnitude 
that led Kormendy (1985b, his figure~3) to advocate a separation of dwarf and normal 
elliptical galaxies at $M_B = -18$ mag.  However, as noted by Graham \&
Guzm\'an (2003), his sample was missing the bridging population near $M_B=-18\pm1$ mag.
Excluding galaxies of this magnitude from Figure~\ref{Fig_abc}a would also
result in two apparently disjoint relations nearly at right angles to each
other. It is therefore easy to understand how one may quickly reach the wrong
conclusion from an incomplete diagram. 
Although, Strom \& Strom (1978, their figure~8; see also Binggeli et al.\
1984) had already revealed that a linear relation exists between magnitude and
central surface brightness from $-18.4 < M_V < -21.6$ mag, spanning the
magnitude gap in Kormendy (1985b). 
Nonetheless, Faber \& Lin (1983) had just observed that three of their six dwarf
elliptical galaxies had near-exponential light profiles, leading them to
speculate that dEs are more closely related to ``exponential systems'' than
(tidally truncated) elliptical galaxies, and Wirth \& Gallagher (1984, see also
Michard 1979) had also just advocated a division between exponential-dwarf 
and $R^{1/4}$-giant elliptical galaxies.  

To further confound matters, 
Kormendy (1985b) had the slope wrong for the distribution of dwarf
elliptical galaxies in his $M$--$\mu_0$ diagram, which had two consequences.
First, the bright end of his dwarf elliptical galaxy 
distribution did not point towards the faint-end 
of his luminous elliptical galaxy distribution, 
and thus there was no suggestion of a 
connection.  This misrepresentation is unfortunately still propagated today
(e.g.\ Tolstoy et al.\ 2009, their figure~1), although Kormendy et al.\ (2009)
have now corrected this.  
Second, two points representing flattened disc galaxies were added at the
bright end of the mis-aligned dwarf elliptical galaxy sequence by Kormendy
(1985b), implying a connection between dwarf elliptical galaxies and 
disc galaxies rather than ordinary elliptical galaxies. 

A decade later, the Astronomy and Astrophysics Review paper by Ferguson \&
Binggeli (1994; their Figure 3) had a big question mark as to ``how'' and
indeed ``if'' dwarf and ordinary elliptical galaxies might connect in this
diagram.  When galaxies spanning the gap in Kormendy's (1985b) analysis were
included by Faber et al.\ (1997, their figure~4c), and shown to follow a trend
consistent with the relation from Strom \& Strom (1978) and Binggeli et al.\
(1984), in which the central surface brightness became fainter with decreasing
galaxy luminosity 
Faber et al.\ (1997) suggested that this behavior in their data 
was spurious and due to limited resolution --- 
such was the belief in a discontinuity separating dwarf and ordinary
elliptical galaxies.
At the same time, Jerjen \& Binggeli (1997) argued exactly the opposite,
suggesting that it was instead the ``core'' galaxies which had been displaced in the
$M$--$\mu_0$ diagram from a linear $M$--$\mu_0$ relation. rather than wrong 
central surface brightness measurements for the faint (non-dwarf) elliptical
galaxies. 
As Graham \& Guzm\'an (2003) and Graham (2004) later explained, 
in terms of a $\sim$0.1 percent central mass deficit relative to the outer
S\'ersic profile in galaxies brighter than $M_B \approx -20.5$ mag, 
Jerjen \& Binggeli (1997) were right, supporting the views 
expressed by Binggeli et al.\ (1984) on a continuity between dwarf elliptical
and ordinary elliptical galaxies across the alleged divide at $M_B \approx
-18$ mag.

\subsubsection{Luminosity-colour relation}
 
Additional support for the dry merging scenario at the high-mass end is
the flattening of the colour-magnitude relation above 0.5-1$\times 10^{11}
M_{\odot}$.
While low luminosity, low S\'ersic index, elliptical galaxies are bluer than
bright elliptical galaxies (e.g.\ de Vaucouleurs 1961; Webb 1964; Sandage
1972; Caldwell \& Bothun 1987), the brightest galaxies have the same colour
as each other.  
This flattening in the colour-magnitude relation was noted by Tremonti et al.\
(2004) and is evident in Baldry et al.\ (2004, their Figure~9), Ferrarese
et al.\ (2006a, their Figure~123), Boselli et al.\ (2008, their Figure~7) and even Metcalfe, Godwin \& Peach (1994).
These observations help alleviate past tension with semi-analytic models that
had predicted a relatively flat colour-magnitude relation for bright
elliptical galaxies (e.g.\ Cole et al.\ 2000).  Previously, based on what was thought
to be a linear colour-magnitude relation, Bernardi et al.\ (2007) had written
that ``if BCGs formed from dry mergers, then BCG progenitors must have been
red for their magnitudes, suggesting that they hosted older stellar
populations than is typical for their luminosities''. However, the flattening
in the colour-magnitude relation 
has since been recognised in yet more data sets (e.g.\ Skelton, Bell \&
Somerville 2009; Jim\'enez et al.\ 2011) 
although it should perhaps be noted
that Skelton et al.\ reported the transition at $M_R=-21$ mag, i.e.\ $\sim$1
mag fainter.

In passing it is noted that the relation between luminosity and supermassive
black hole mass (Marconi \& Hunt 2003; McLure \& Dunlop 2004) was found to be,
after several refinements by Graham (2007), a linear one-to-one relation for
black hole masses predominantly greater than $10^8 M_{\odot}$ --- consistent
with the concept of dry galaxy merging at this high-mass end.

\subsubsection{Dynamics}\label{Sec_Faber}

From a sample of 13 early-type galaxies, plus one spiral galaxy, Minkowski
(1962) noted that a ``correlation between velocity dispersion and [luminosity]
exists, but it is poor''. He wrote that ``it seems important to extend the
observations to more objects, especially at low and medium absolute
magnitudes''. This was done by Morton \& Chevalier (1973) who noted the same
``continuous distribution of dispersions from 60 km/s for M32 to 490 km/s for
M87'' but also did not attempt to quantify this trend.
It was Faber \& Jackson (1976) who, with improved data and a larger sample of 25 galaxies, were
the first to quantify Minkowski's relation and discovered that $L\propto 
\sigma^4$ for their data set. This result has proved extremely popular and is known as the 
Faber-Jackson relation. 
Not long after this, Schechter (1980) 
and Malumuth \& Kirshner (1981) 
revealed that the luminous elliptical galaxies followed a relation with an
exponent of $\sim$5 rather than 4.  At the same time, Tonry (1981) revealed
that expanding the sample to include more faint elliptical galaxies results in
an exponent of $\sim$3.  This led Binney (1982) to write that ``probably the
correlation cannot be adequately fitted by a single power law over the full
range of absolute magnitudes'' and Farouki et al.\ (1983) wrote that ``the
data suggests the presence of curvature in the $L-\sigma$ relation''.  Davies
et al.\ (1983), and later Held et al.\ (1992), revealed that the dwarf
elliptical galaxies followed a relation with an exponent of $\sim$2, which
explains why Tonry (1981) had found a slope of $\sim$3 when including dwarf and
ordinary elliptical galaxies.  The relation found by Davies et al., with a slope of $\sim$2, has
recently been observed by de Rijcke et al.\ (2005) and the curved or possibly
broken $L-\sigma$ distribution has been interpreted by Matkovi\'c \& Guzm\'an
(2005) as a change in slope at $-20.5$ B-mag (see also Evstigneeva et al.\
2007) in agreement with Davies et al.\ (1983). 

In spite of all the above work, there is a huge body of literature today 
which appears unaware that the $L-\sigma$ relation is curved or broken. 
Simulations of galaxies which succeed in producing the linear Faber-Jackson
relation, $L\propto \sigma^4$, have actually failed to produce the full
distribution of dynamics seen in real elliptical galaxies as a function of magnitude.

\section{Disc Galaxy Light Profiles}\label{Sec_Discs}

Due to contractual agreements with Springer Publishing, the full version of
this paper can not be posted to astro-ph.  
All of Section~\ref{Sec_Discs}, plus the full version of section~\ref{L-Mu0}, 
will be available from the Springer Publication.

\subsection{The Bulge-Disc Decomposition}\label{Sec_BD}

\subsection{Dust and inclination corrections}\label{sec_dust}

\subsection{Pseudobulges}\label{sec_pseudo}

\subsubsection{S\'ersic index}

\subsubsection{Rotation}

\subsubsection{Ages}

\subsubsection{Scaling relations}

\subsection{Bulgeless galaxies}\label{sec_less}

\subsection{Barred galaxies}\label{sec_bar}

\vspace{1cm}

\section{Summary}

We have reviewed the progress over the last century in modelling the
distribution of stars in elliptical galaxies, plus the bulges of lenticular
and spiral galaxies and their surrounding discs.  A number of nearly forgotten
or poorly recognised references have been identified.  The universality, or at
least versatility, of S\'ersic's $R^{1/n}$ model to describe bulges
(section~\ref{Sec_BD}) and elliptical galaxies (section~\ref{Sec_body}) 
extends to the stellar halos of cD galaxies (section~\ref{Sec_cD}) and 
simulated dark matter halos (section~\ref{Sec_dark}).

Dwarf and ordinary elliptical galaxies were shown in section~\ref{Sec_Line} to
be united by two continuous linear relations between absolute magnitude and a)
the stellar concentration quantified through S\'ersic's (1963) $R^{1/n}$ shape
parameter (section~\ref{Sec_M-n}), and b) the central surface brightness,
which is also related to the central density (section~\ref{Sec_M-mu0}).  As
discussed in section~\ref{Sec_bent}, a break in the latter relation at
$M_B\approx -20.5$ mag signals the onset of partially depleted cores relative
to the outer S\'ersic profile in luminous elliptical galaxies.  
Additional scaling relations are also noted to
show a change in character at this magnitude, 
which may denote the onset of dry galaxy merging.

The identification of depleted galaxy cores and excess nuclear light relative
to the outer S\'ersic profile was discussed in sections~\ref{Sec_cS},
\ref{Sec_def} and \ref{Sec_Enuc}.  After accounting for these features, it was
revealed how the above two linear relations result in curved scaling relations
involving effective half light radii and effective surface brightness
(section~\ref{sec_curved}).  Specifically, the $M$--$R_{\rm e}$,
$M$--$\mu_{\rm e}$, $M$--$\langle \mu \rangle_{\rm e}$, $\mu_{\rm e}$--$R_{\rm
  e}$, $\langle \mu \rangle_{\rm e}$--$R_{\rm e}$ and $n$--$R_{\rm e}$
relations are non-linear.  These continuous curved relations exist because
elliptical galaxies do not have a universal profile shape, such as an
$R^{1/4}$ profile, but instead a range of profile shapes that vary smoothly
with absolute magnitude.
Without an appreciation of the origin of these curved relations, they had in
the past been heralded as evidence for a dichotomy between faint and bright
elliptical galaxies.  Numerical simulations and semi-numerical models which
try to reproduce the full elliptical galaxy sequence must be able to reproduce
these non-linear relations.  This will likely require physical processes which
work in tandem, albeit to different degrees over different mass ranges, to
produce a continuum of galaxy properties that scale with mass while adhering
to the linear $M$--$n$ and $M$--$\mu_0$ relations (subject to core-formation).

[snip]

The upcoming 
2.6 m VLT Survey Telescope (VST, Arnaboldi et al.\ 1998; Capaccioli et al.\ 2005), plus the 
4$\times$1.8 m Pan-STARRS array (Kaiser et al.\ 2002), the 
4 m Visible and Infrared Survey Telescope for Astronomy (VISTA,
Emerson et al.\ 2004) and the 8.4 m Large Synoptic Survey Telescope
(LSST, Tyson 2001) are expected to deliver sub-arcsecond, deep and wide
field-of-view imaging covering thousands of resolvable galaxies. 
By pushing down the luminosity function into the dwarf galaxy regime, and
through the application of improved galaxy parameterisation methods which
allow for structural non-homology and the 2- or 3-component nature of disc
galaxies, {\it both} statistical and systematic errors will be reduced.  This will
undoubtedly provide improved constraints on galaxy scaling relations and, in
turn, a fuller understanding of galaxy evolution.

\acknowledgments

This research was supported under the Australian Research Council's Discovery
Projects funding scheme (DP110103509). 
This research has made use of NASA's Astrophysics Data System (ADS)
Bibliographic Services and the NASA/IPAC Extragalactic Database (NED).

\clearpage

\end{document}